\documentclass[useAMS,usenatbib]{mn2e}
\usepackage{graphics}
\usepackage{graphicx}
\usepackage{epsf}
\usepackage{url}

\def\bk{{\bf k}}

\def\br{{\bf r}}
\def\LCDM{$\Lambda$CDM}
\def\hMsun{{\ }h^{-1}{\,}{\rm M_{\odot}}}
\def\hMpc{{\ }h^{-1}{\,}{\rm Mpc}}
\def\ihMpc{{\ }h{\,}{\rm Mpc}^{-1}}
\def\VOBOZ{{\scshape voboz}}
\def\FFTLog{FFTLog}
\newcommand{\yeq}{
  \begin{figure}
    \begin{center}
     \leavevmode
      \epsfxsize=\columnwidth   
      \epsfbox{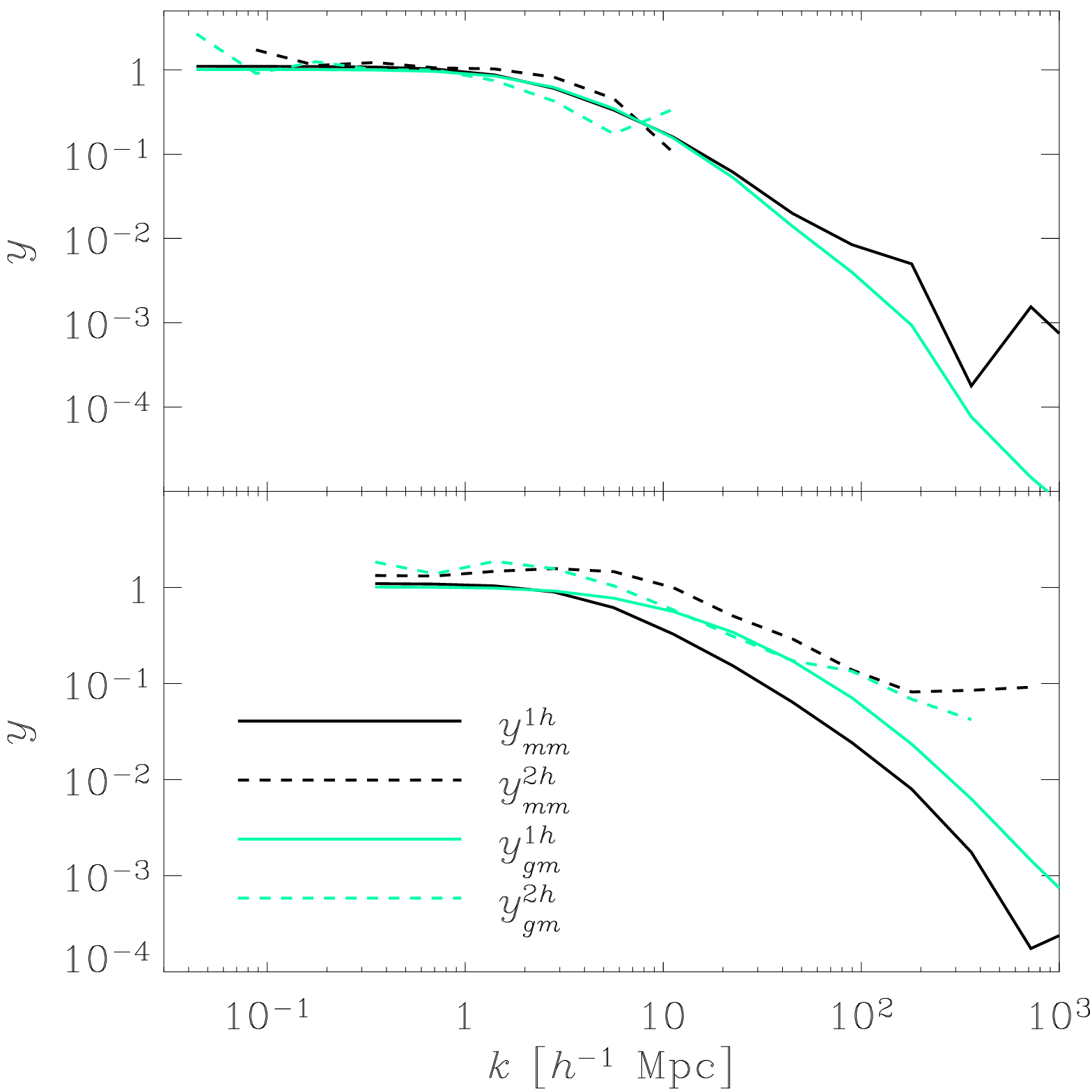}
    \end{center}
    \caption[1]{ \small For two different halo populations, comparisons of
    the average Fourier-transformed density profiles $y_{gm}^{1h}$,
    $y_{gm}^{2h}$, $y_{mm}^{1h}$, and $y_{mm}^{2h}$, as measured from
    $N$-body simulations described by \citet{nhg}.  The mass threshold
    defining the top panel's halo sample is $3\times 10^{13}\hMpc$; the
    bottom panel's threshold is smaller by a factor of 512.  The $1h$
    profiles $y_{gm}^{1h}$ and $y_{mm}^{1h}$ were obtained from
    $P_{gm}^{1h}$ and $P_{mm}^{1h}$ (measured by Fourier-transforming
    $\xi_{gm}^{1h}$ and $\xi_{mm}^{1h}$) using eqs.\ (\ref{pgm1h}) and
    (\ref{pmm1h}).  The $2h$ profiles $y_{gm}^{2h}$ and $y_{mm}^{2h}$ were
    measured from $P_{gm}^{2h}$, $P_{mm}^{2h}$, and $P$ using
    eqs. (\ref{pgm2h}) and (\ref{pmm2h}).  These latter two equations
    require an estimate of the large-scale bias $b_0$ for each halo sample.
    For the top panel, we fitted $b_0 = 1$ by hand, and for the bottom
    panel, we used the value $b_0 = 0.75$ used for Fig.\ \ref{simb}, which
    is a plot of the bias from the same halo sample.
      \label{yeq}
    }
  \end{figure}
}

\newcommand{\nfwhp}{
  \begin{figure}
    \begin{center}
     \leavevmode
      \epsfxsize=\columnwidth   
      \epsfbox{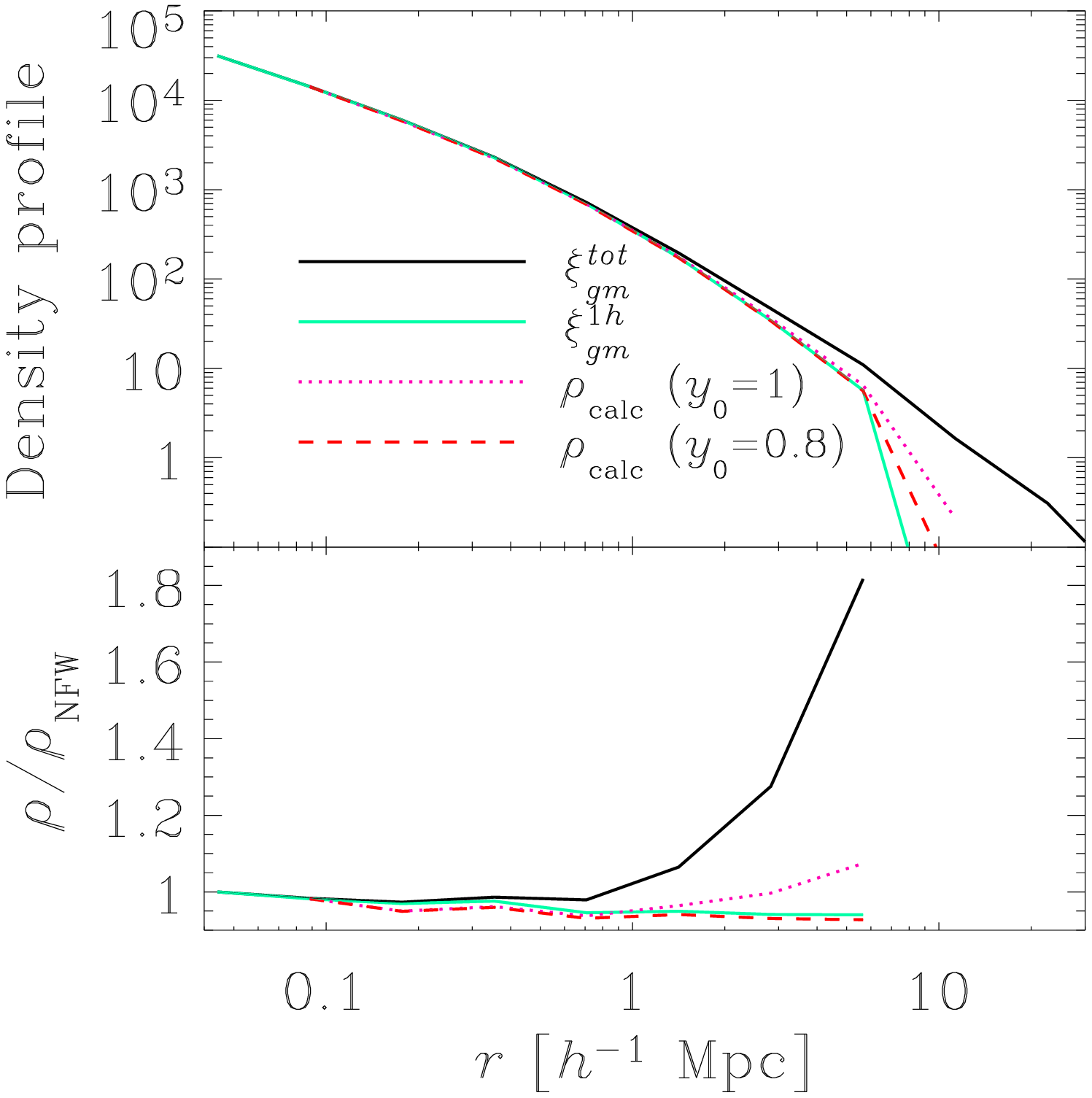}
    \end{center}
    \caption[1]{ \small An NFW density profile, with a scale radius of
      $1\hMpc$ and truncated at $8\hMpc$, along with various attempts to
      recover it from a mock catalog with haloes with the profile.  In the
      top panel, the solid black curve shows the total $1+\xi_{gm}$, and
      the solid grey (online, green) curve shows the one-halo
      $1+\xi_{gm}^{1h}$.  The dashed grey (online, magenta) curve shows our
      attempt to isolate the $1h$ term, assuming $y=1$ in the
      smallest-wavenumber bin, and the dotted grey curve shows the same
      assuming $y=0.8$ in the smallest-wavenumber bin.  The bottom panel
      shows the ratio of the respective curves in the top panel with the
      true NFW profile.  The curves disappear for $r>8\hMpc$ (actually,
      $6\hMpc$, the centre of the bin which goes up to $8\hMpc$) because
      the input profile is truncated there.
      \label{nfwhp}
    }
  \end{figure}
}

\newcommand{\scdr} {
  \begin{figure}
    \begin{center}
     \leavevmode
      \epsfxsize=\columnwidth   
      \epsfbox{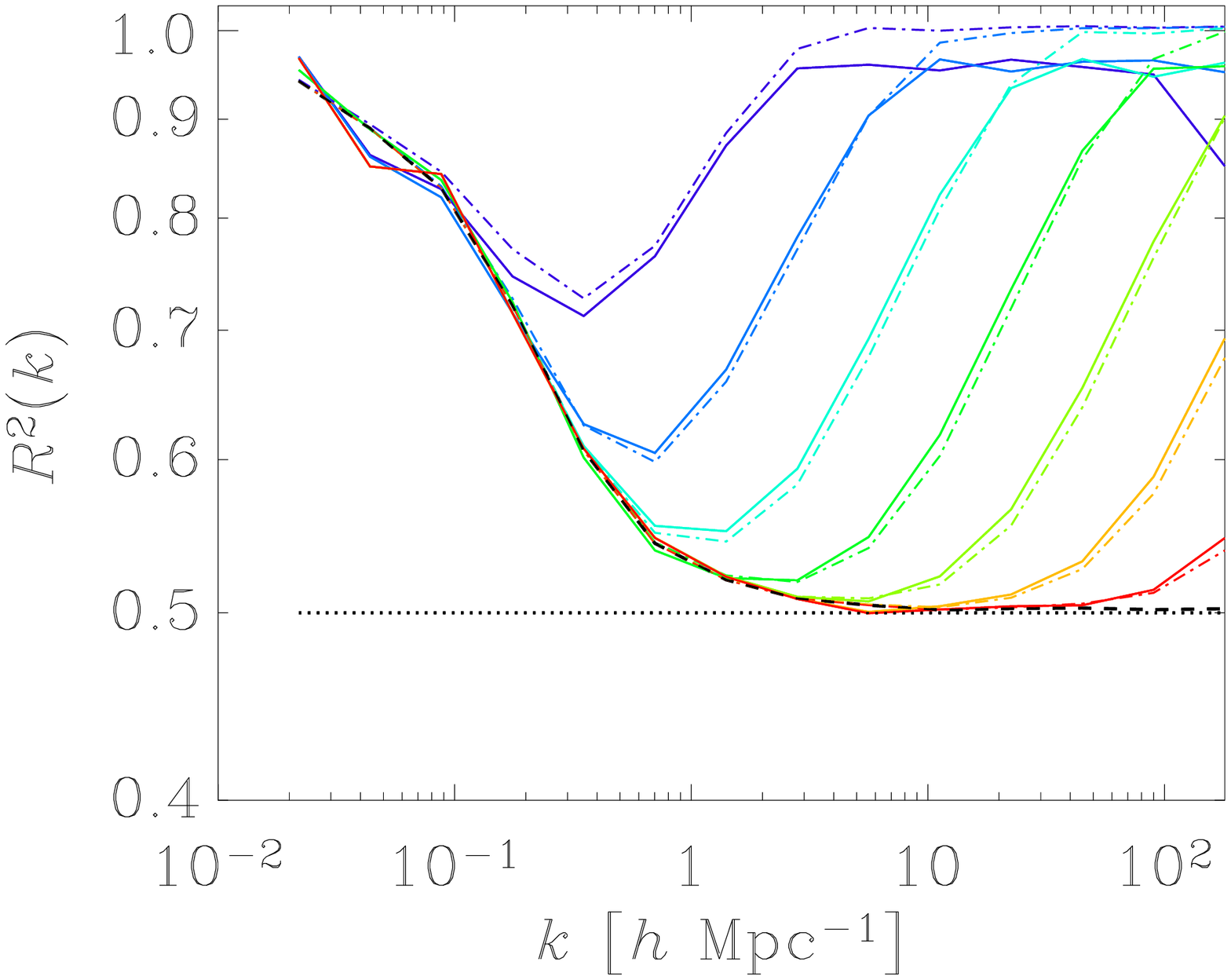}
    \end{center}
    \caption[1]{ \small Testing eqs.\ (\ref{r2ay}) and (\ref{r2a}) for
      $R^2(k)$.  Density profiles $\rho(r)\sim r^{-2}$ were put around
      galaxies such that the central densities were held fixed, but the
      haloes grew in radius, proportional to their masses.  The dotted line
      is at $1/\mu=0.5$, and the dashed curve is approximation (\ref{r2a}).
      The solid curves depict $R^2$ for different halo radii per particle.
      The halo radius per particle varies by factors of $\sqrt{10}$ from
      $10^{-5.5}$ (rightmost upturn) to $10^{-2.5}\hMpc$ (leftmost upturn).
      For the set of haloes constructed with $10^{-2.5}\hMpc$ per particle,
      the minimum and maximum halo radii were 0.36 and $16\hMpc$.  The
      dot-dashed curves show the approximations given by eqn.\
      (\ref{r2ay}), with $(y_{gm}/y_{mm})^2$ calculated analytically.
      \label{scdr}
    }
  \end{figure}
}

\newcommand{\simr}{
    \begin{figure*}
    \begin{minipage}{175mm}
    \begin{center}
    \leavevmode
    \epsfxsize=\columnwidth   
    \epsfbox{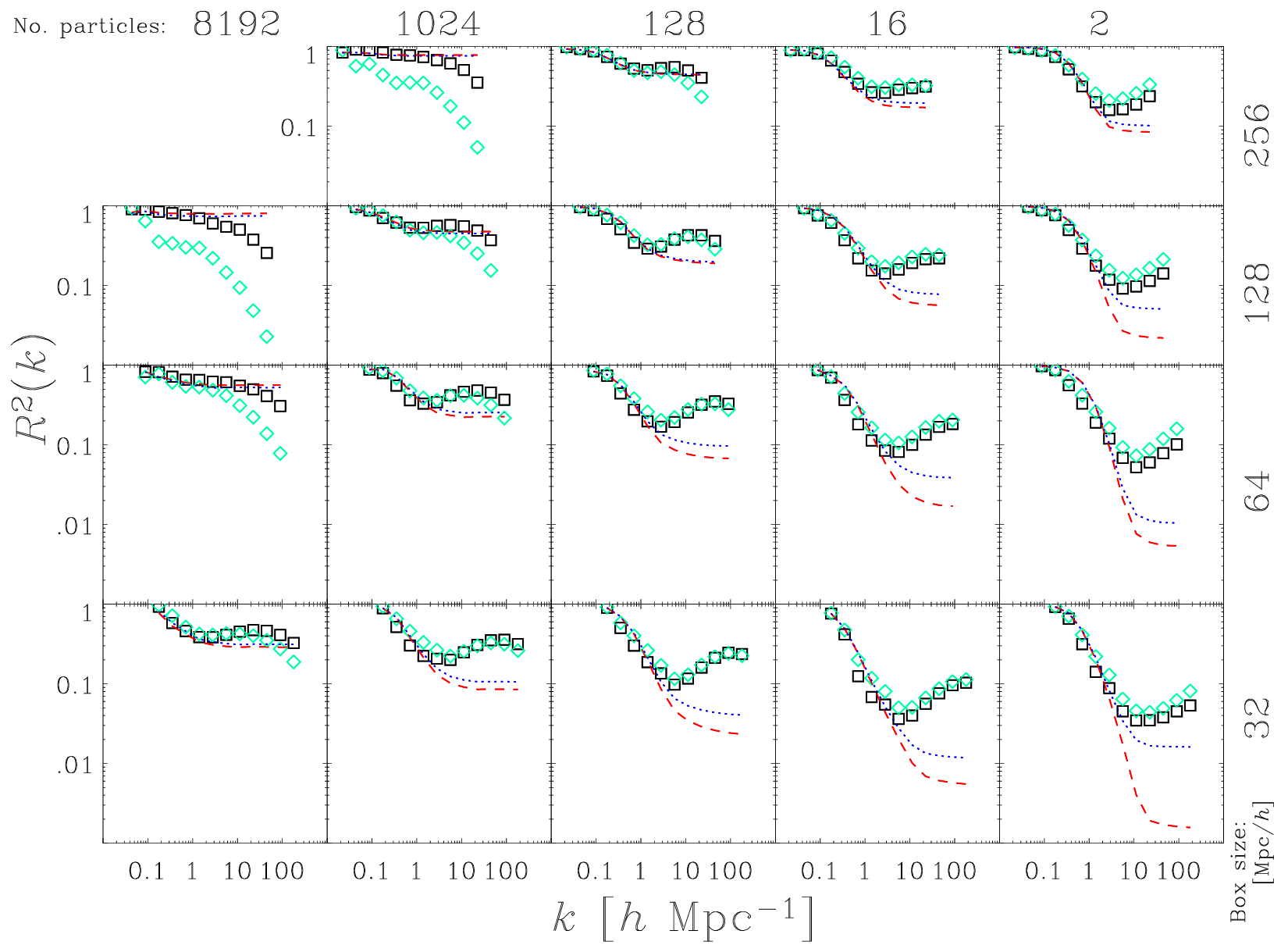}
    \end{center}
    \caption[1]{ \small Testing approximation (\ref{r2a}) for the
      galaxy-matter cross-correlation coefficient $R^2(k)$ in actual
      simulations, for different halo mass cut-offs and box sizes.  From
      left to right, the halo mass (given in particle number) cut-off
      decreases by a factor of eight, the same factor by which the particle
      mass decreases from top to bottom with box size.  Thus, the physical
      halo mass is the same along diagonals starting from the bottom-left
      and going up and right.  The squares show $R^2$ calculated using only
      particles in haloes as identified by {\VOBOZ}, while the grey
      (online, green) diamonds show $R^2$ using all particles in the
      simulation.  The dashed (online, red) lines show approximation
      (\ref{r2a}), with the dimensionless mean-square halo mass $\mu$
      calculated from the {\VOBOZ} halo masses.  For the dotted (online,
      blue) lines, a smaller $\mu^\prime$ was used, calculated by
      identifying clusters of haloes (i.e.\ a halo and its subhaloes) such
      that each subhalo is within the half-mass radius of a parent halo,
      and then distributing the mass of the parent halo equally among the
      haloes and subhaloes.  We fixed the large-scale bias $b_0$ by
      requiring that all of the curves line up in the largest-scale,
      lowest-wavenumber bin.
      \label{simr}
    }
    \end{minipage}
    \end{figure*}
}

\newcommand{\difmu}{
    \begin{figure}
    \begin{center}
    \leavevmode
    \epsfxsize=\columnwidth   
    \epsfbox{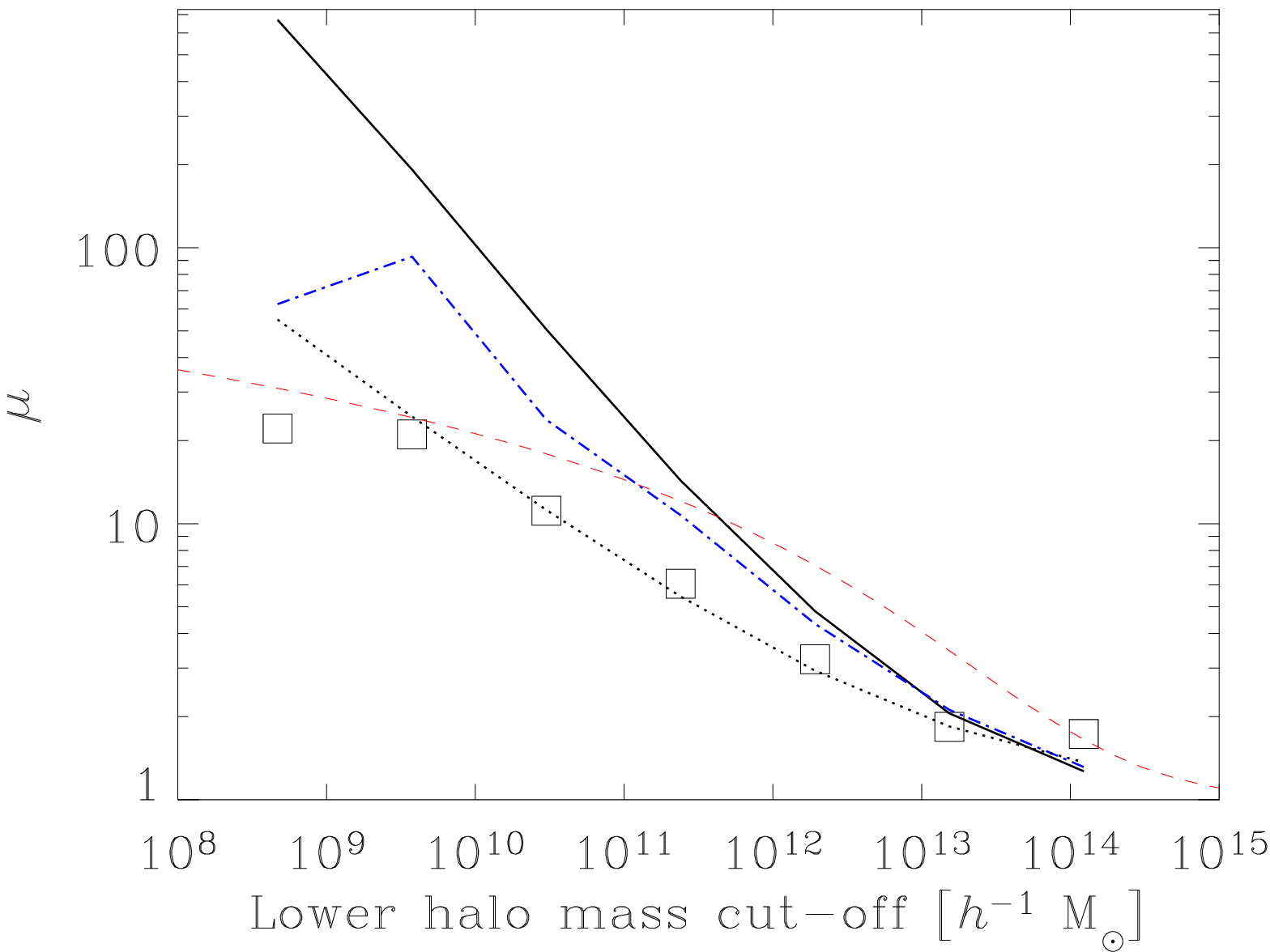}
    \end{center}
    \caption[1]{ \small An illustration of different estimates of the
      dimensionless mean-square halo mass $\mu$ from four {\LCDM} $N$-body
      simulations.  The abcissa is the lower mass threshold used to define
      each halo sample.  The curves are averages of the results from the
      four nested simulations.  The positions of the squares show $\mu_{\rm
      meas} = 1/R^2_{\rm min}$, where $R^2_{\rm min}$ is the lowest value
      $R^2(k)$ takes for $k<10\ihMpc$, as measured from the $R^2(k)$ curves
      in Fig.\ \ref{simr}.  The solid curve shows $\mu$ as calculated from
      the {\VOBOZ} mass distributions.  The (online, blue) dashdotted curve
      shows $\mu^\prime$, for which we reduced the dispersion by assigning
      all of the mass in clusters equally among their subhaloes.  The
      dashed curve shows $\mu$ as calculated from the \citet{st99} mass
      function, using no upper mass threshold.  The dotted line shows a
      crude fit to $\mu$, given in eq.\ (\ref{mumeas}).  The height of the
      squares roughly indicates the scatter, which was similar in all
      curves.  For very low and high mass cut-offs, the averages contained
      fewer than four samples because the cut-offs could not be used in all
      simulations.
      \label{difmu}
    }
    \end{figure}
}

\newcommand{\simb}{
    \begin{figure}
    \begin{center}
    \leavevmode
    \epsfxsize=\columnwidth   
    \epsfbox{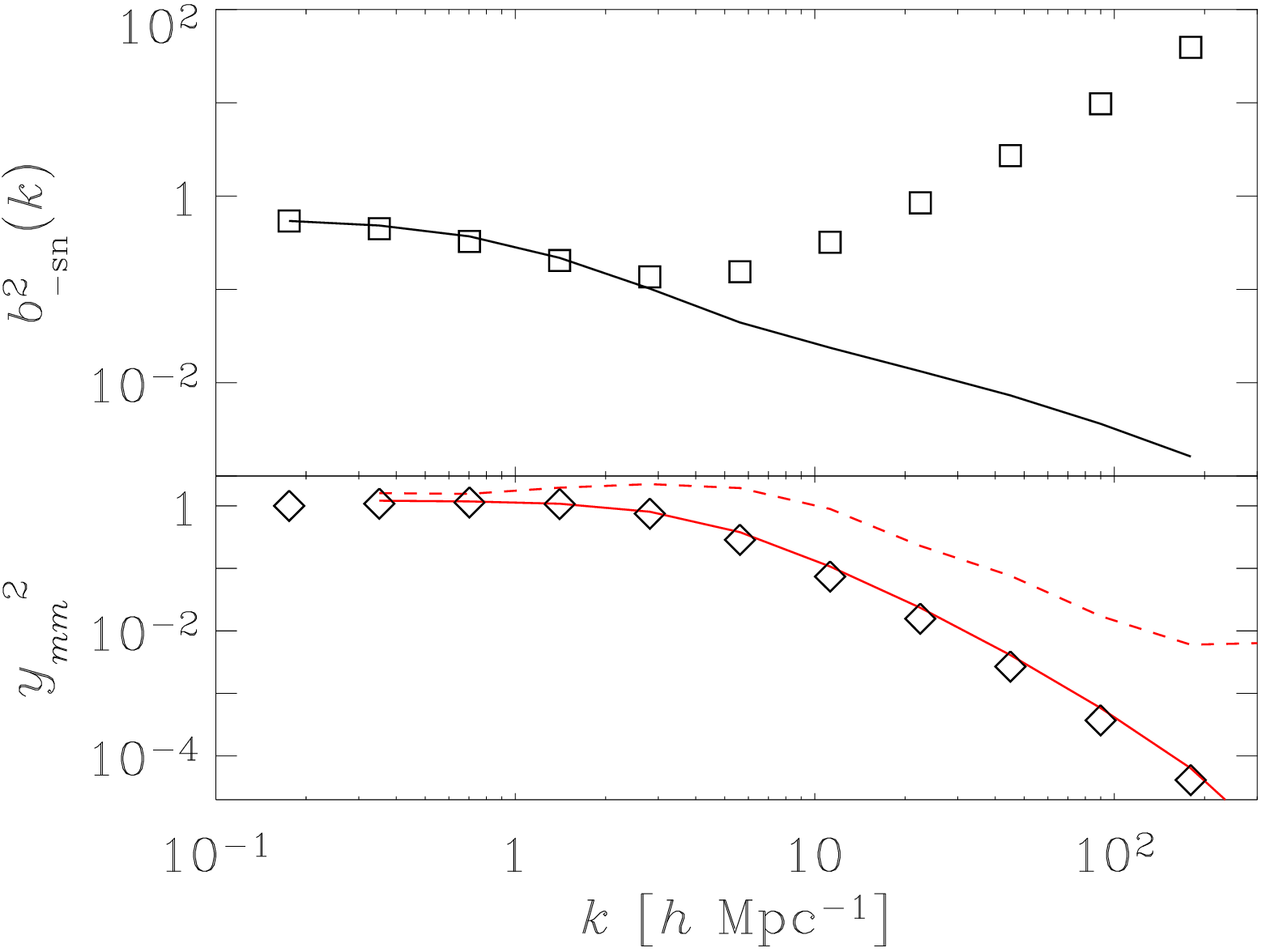}
    \end{center}
    \caption[1]{ \small In the top panel, squares show the squared bias
    $b_{\rm -sn}^2(k)$ (excluding the shot noise from $P_{gg}$), and the
    curve shows the result of the galaxy-halo model, eq.\ (\ref{b}),
    assuming that haloes are `infinitesimal nuggets', i.e.\ that
    $y_{mm}(k)=1$.  The value of the large-scale bias $b_0$ has been
    adjusted so that the squares and the curve in the top panel agree at
    the largest scale plotted.  In the bottom panel, the diamonds show the
    ratio of the curve in the top panel to the squares; this ratio measures
    the mean-square average Fourier-transformed halo density profile
    $y_{mm}(k)^2$, assuming that $y_{mm}^{1h} = y_{mm}^{2h}$, as in eq.\
    (\ref{b}).  For comparison, the grey (online, red) curves show
    $y_{mm}^{1h}(k)^2$ (solid) and $y_{mm}^{2h}(k)^2$ (dashed) from the
    same halo sample.  (The square roots of these two latter curves appear
    in Fig.\ \ref{yeq}.)
    \label{simb}
    }
    \end{figure}
}

\newcommand{\idithp}{
  \begin{figure}
    \begin{center}
     \leavevmode
      \epsfxsize=\columnwidth   
      \epsfbox{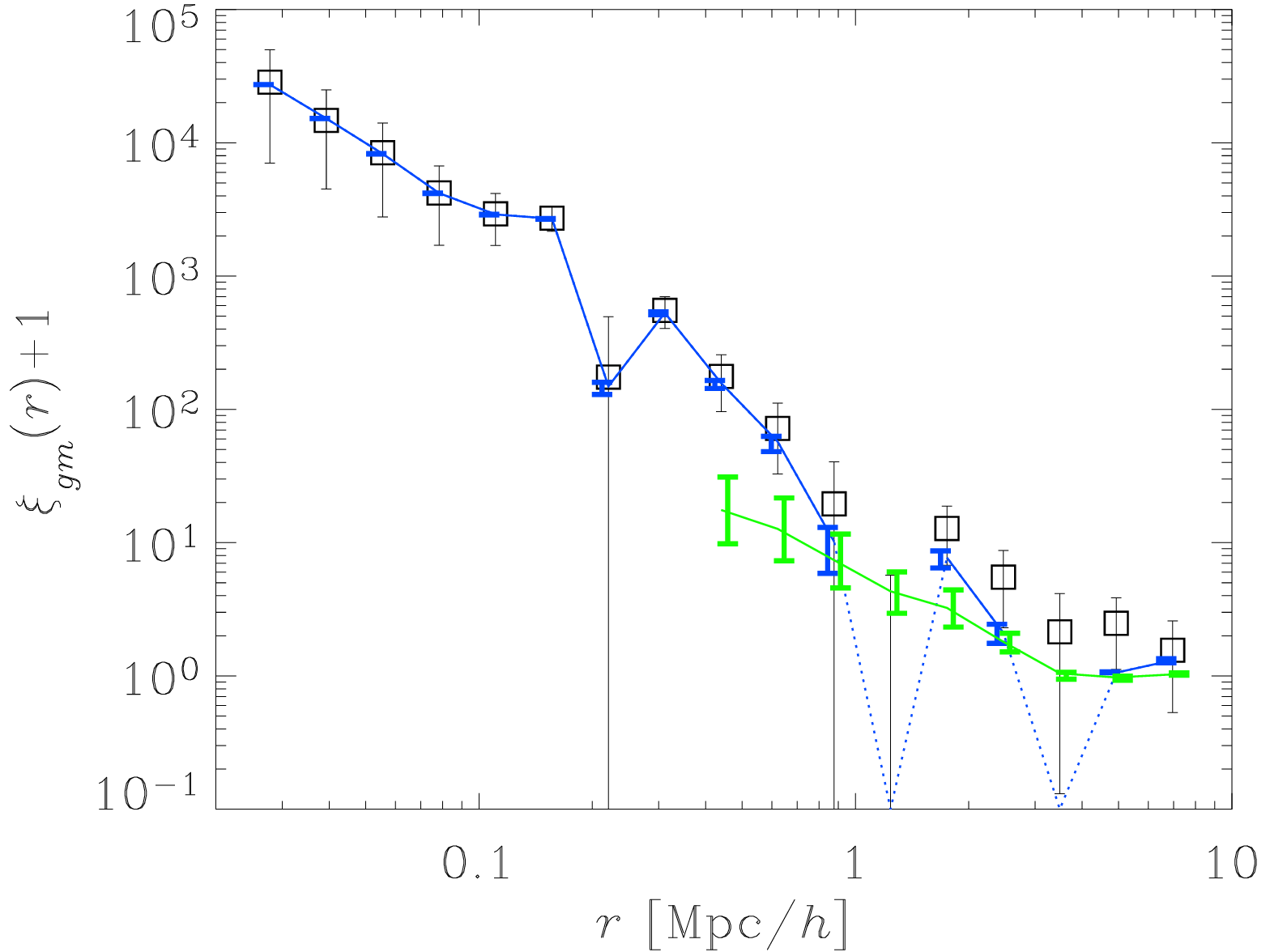}
    \end{center}
    \caption[1]{ \small An attempt to recover the average overdensity
      profile of haloes around galaxies in a volume-limited sample of
      luminous SDSS galaxies from the galaxy-matter correlation function,
      using the galaxy correlation function.  The squares with error bars
      are the measurement by \citet{sheldon} of $\xi_{gm}$ (plus 1), and
      the dark grey (online, blue) solid curve is the effective one-halo
      term $\xi_{gm}^{1h}$, or $\bar\delta (r)$ (plus 1). This curve
      becomes dotted, and goes to the bottom of the plot, where
      $\xi_{gm}^{1h}+1 < 0$.  The thick error bars around data points in
      the one-halo curve show the results of varying the large-scale bias
      $b_0$ by factors of 2 of the canonical value, $b_0 = 1.12$, in both
      directions.  The light grey (online, green) curve is the two-halo
      term, $\xi_{gm}^{2h}+1$, again with thick error bars showing the
      fluctuation it experiences as $b_0$ changes by factors of 2.
      \label{idithp}
    }
  \end{figure}
}

\newcommand{\iditr}{
  \begin{figure}
   \begin{center}
    \leavevmode
    \epsfxsize=\columnwidth 
    \epsfbox{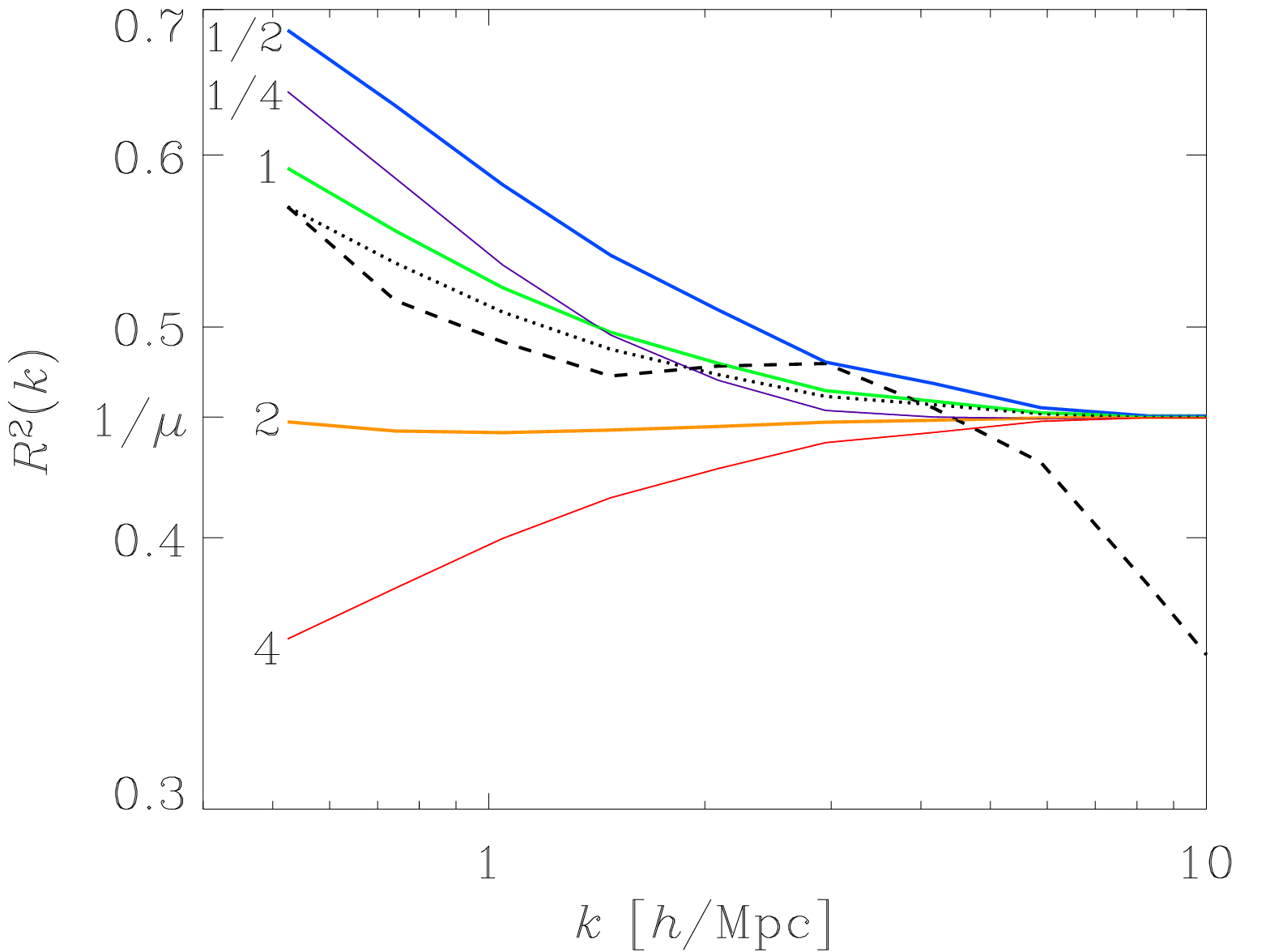}
   \end{center}
   \caption[1]{ \small An illustration of how the galaxy-matter
  cross-correlation coefficient $R^2(k)$, calculated using eq.\
  (\ref{r2a}), depends on the large-scale bias parameter $b_0$.  The galaxy
  spectrum $P$ is taken from from a volume-limited sample of luminous SDSS
  galaxies \citep{z02}.  The dimensionless mean-square halo mass $\mu$ is
  fixed at $2.2$.  Solid curves are labelled according to the values of
  $b_0$ used to calculate them.  For realistic halo samples, $b_0$ should
  seldom stray outside the range $1/2<b_0<2$; in extreme cases, it might
  venture out to $1/4$ or $4$.  The curve at $b_0=1/2$ is nearly identical
  to the highest-possible $R^2(k)$ for each $k$, which uses
  $b_0=1/\mu\approx 0.45$.  The dashed curve shows $R^2$ measured from a
  simulation described in the text.  The dashed curve shows $R^2$ from eq.\
  (\ref{r2a}), using $b_0 = 1.12$, the value of $b_0$ which gives a curve
  matching the dashed curve in the lowest-wavenumber bin, at $k\approx 0.5
  \ihMpc$.
   \label{iditr}
  }
 \end{figure}
}

\newcommand{\iditb}{
  \begin{figure}
    \begin{center}
     \leavevmode
      \epsfxsize=\columnwidth   
      \epsfbox{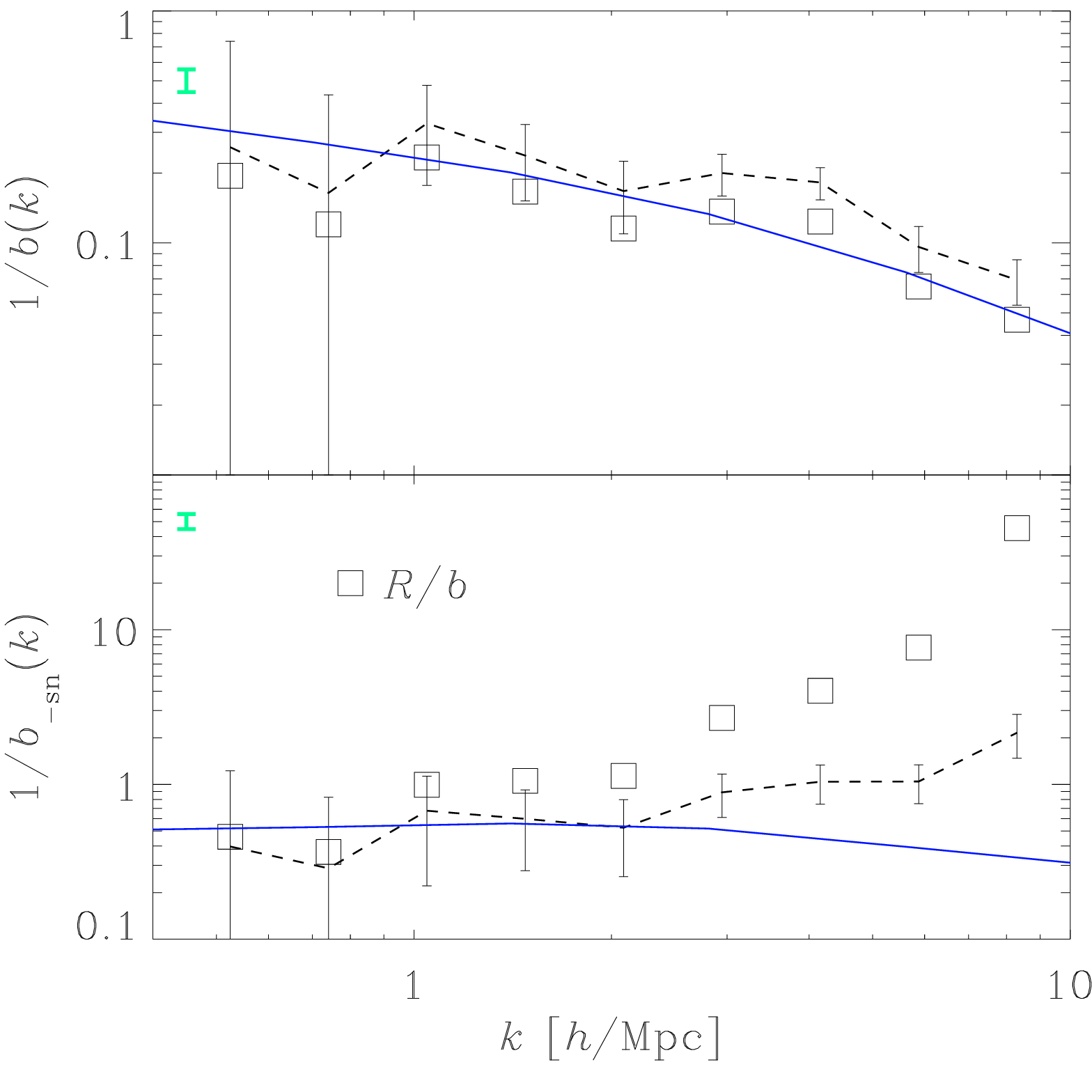}
    \end{center}
    \caption[1]{ \small An attempt to measure the bias $b(k)$ (with the
      shot noise included in $P_{gg}$) and $b_{\rm -sn}(k)$ (without the
      shot noise in $P_{gg}$) of a volume-limited sample of luminous SDSS
      galaxies.  Because the error bars are larger in $P_{gm}$, we put it
      in the numerator, showing $1/b$ instead of $b$.  The squares show the
      cross-bias $R(k)/b(k) = P_{gm}(k)/P_{gg}(k)$, measured from
      $\xi_{gm}(r)$ \citep{sheldon} and $\xi_{gg}(r)$ \citep{z02}.  The
      dashed curves show $1/b$ as calculated by dividing $R/b$ by $R$ as
      calculated from eq.\ (\ref{r2a}), and as shown in Fig.\ \ref{iditr}.
      For the dashed curves, we used a dimensionless mean-square halo mass
      $\mu=2.2$ and a large-scale bias $b_0=1.12$.  The isolated, thick,
      grey (online, green) error bars in the upper-left corners of each
      panel show the largest fluctuation (in the smallest-wavenumber bin)
      in $1/b$ and $1/b_{\rm -sn}$ and if $b_0$ varies by factors of two in
      both directions from $1.12$.  The thin, black error bars are the
      observational error bars, propagated through the analysis.  The
      (online, blue) solid curves are $1/b$ measured from a halo catalog in
      a simulation; the halo catalog has a lower mass cut-off giving the
      same number density as the observed galaxy catalog.
      \label{iditb}
    }
  \end{figure}
}

\begin{document}
\title[A galaxy-halo model of large-scale structure]
	{A galaxy-halo model of large-scale structure}

\author[Mark C.\ Neyrinck, Andrew J.\ S.\ Hamilton, and Nickolay Y.\ Gnedin]
{Mark C.\ Neyrinck$^{1,2}$, Andrew J.\ S.\ Hamilton$^{1,2}$, and Nickolay Y.\ Gnedin$^{2,3}$ \\
  $^{1}$JILA, University of Colorado, Boulder, CO 80309\\
  $^{2}$Department of Astrophysical and Planetary Sciences, University of Colorado, Boulder, CO 80309 \\
  $^{3}$Center for Astrophysics and Space Astronomy, University of Colorado, Boulder, CO 80309\\
  email: {\tt Mark.Neyrinck@colorado.edu}}
\date{2005 Jun 9}

\bibliographystyle{mnras}
	 
\maketitle
  
\begin{abstract}
  We present a new, galaxy-halo model of large-scale structure, in which
  the galaxies entering a given sample are the fundamental objects.  Haloes
  attach to galaxies, in contrast to the standard halo model, in which
  galaxies attach to haloes.  The galaxy-halo model pertains mainly to the
  relationships between the power spectra of galaxies and mass, and their
  cross-power spectrum.  With surprisingly little input, an
  intuition-aiding approximation to the galaxy-matter cross-correlation
  coefficient $R(k)$ emerges, in terms of the halo mass dispersion.  This
  approximation seems valid to mildly non-linear scales ($k \la 3\ihMpc$),
  allowing measurement of the bias and the matter power spectrum from
  measurements of the galaxy and galaxy-matter power spectra (or
  correlation functions).  This is especially relevant given the recent
  advances in precision in measurements of the galaxy-matter correlation
  function from weak gravitational lensing.  The galaxy-halo model also
  addresses the issue of interpreting the galaxy-matter correlation
  function as an average halo density profile, and provides a simple
  description of galaxy bias as a function of scale.
\end{abstract}

\begin {keywords}
  large-scale structure of Universe -- galaxies: haloes -- cosmology:
  theory -- dark matter -- galaxies: formation -- gravitational lensing.
\end {keywords}

\section{Introduction}
The halo model of large-scale structure has been quite successful in
interpreting observations of galaxy and mass clustering.  In the halo
model, the dark matter in the Universe consists entirely of virialized
clumps called haloes.  The assumption is made that galaxies can only form
within these haloes, with the number of galaxies per halo depending
(primarily) on the mass of the halo, according to a halo occupation
distribution (HOD).

Some of the inputs to the halo model arise from pure theory, such as the
linear power spectrum.  Others come from theory with some calibration with
simulations, such as the mass spectrum and bias of the haloes.  Yet others
are entirely empirical, such as the HOD, and halo density profiles.  In
fact, although ideas related to the halo model have existed for decades
\citep{ney,bert}, it was not until a universal halo density profile was
discovered from simulations \citep{nfw97} that the halo model began to
really develop \citep{ps,s00,mf,sshj,bw}.  For a review, see
\citet{cs}. There are some things about the halo model which unsettle us
somewhat; for instance, the cosmic web seen in observations and numerical
simulations is not directly explained in the halo model as it currently
stands.  However, the halo model has matured to the degree that it seems to
be able to help constrain cosmological parameters \citep{s05a,s05b}.  As
observations and simulations improve, it is likely that the halo model will
evolve to match observables from them arbitrarily well as theoretical
ingredients are added to it, without fundamental conceptual changes.

In this paper, we approach the problem of interpreting large-scale
structure observations with a different philosophy, investigating how much
can be learned with the observations themselves, with as little theoretical
input as possible.  The galaxy power spectrum $P_{gg}$ is taken as the
fundamental quantity, even though it is not known how to produce it
theoretically at present.  Haloes (or subhaloes; they are not
differentiated in the model) are attached to galaxies, not the other way
around.  The galaxy-matter power spectrum $P_{gm}$ and the matter power
spectrum $P_{mm}$ (or, more accurately, their two-halo terms) are then
simple convolutions of $P_{gg}$ with average halo density profiles.  Some
added information about large-scale bias and the halo mass dispersion
yields a surprising amount of information about the galaxy-matter
cross-correlation coefficient, the bias, and the cross-bias.

We emphasize that this galaxy-halo model is not meant to be a competitor to
the standard halo model, which could be called the `largest virialized'
halo model, since each halo in the standard halo model is the largest
possible virialized structure.  The galaxy-halo model as presented in this
paper is not a formalism which may be used to compare to all conceivable
observations, although it may evolve to be more encompassing in the future.
It is currently restricted because it has few ingredients, and it is not
known how to obtain the main ingredient $P_{gg}$ a priori.  In fact, we are
surprised that because it is simpler, the galaxy-halo model was not
developed before the halo model.  The above criticism of the halo
model (that it does not produce the cosmic web) applies to the galaxy-halo
model, too.  However, the simplicity of the galaxy-halo model has benefits;
it allows for some intuitive insights, and for interpretation of some
observations with few assumptions.

\section{Ratios of clustering statistics}\label{powspec}
It can be useful to form ratios of two-point statistics \citep[the power
spectrum and the correlation function; see e.g.][]{peebles,h05} which
measure the clustering of matter and galaxies.  In Fourier space, the
(squared) bias is $b^2(k) \equiv P_{gg}(k)/P_{mm}(k)$, the ratio of the
galaxy power spectrum to the matter power spectrum.  The (squared)
galaxy-matter cross-correlation coefficient is $R^2(k) \equiv
P_{gm}(k)^2/[P_{gg}(k)P_{mm}(k)]$, where $P_{gm}$ is the galaxy-matter
cross power spectrum.  With the advent of galaxy-matter observations from
weak lensing, the `cross-bias' has also been used: $b/R \equiv
P_{gg}(k)/P_{gm}(k)$.

The same ratios may also be defined in real space, using correlation
functions instead of power spectra; recent observations
\citep[e.g.][]{sheldon} have tended to favor the real-space description.
There are at least two good reasons for this: $\xi_{gm}$, and not $P_{gm}$,
is more directly measured from weak-lensing observations; and also, $\xi$
is easier to understand intuitively and visually.

However, the theoretically preferred representation is in Fourier space,
for several reasons.  Most relevantly for the present paper, the Schwarz
inequality imposes a mathematical constraint on the cross-correlation
coefficient $R(k)$ when expressed in Fourier space.  The Schwarz inequality
requires that $\langle|\delta_m(\bk)\delta_g(-\bk)|\rangle^2 \le
\langle|\delta_g(\bk)|^2\rangle\langle|\delta_m(\bk)|^2\rangle$.  {\it If
and only if} the galaxy and matter power spectra include the shot noise,
this gives
\begin{equation}
R^2(k)=\frac{P_{gm}(k)^2}{P_{gg}(k)P_{mm}(k)} \le 1.
  \label{r2schw}
\end{equation}
The shot noise in $P_{mm}$ is negligible because dark-matter particles are
practically infinitesimal on all astrophysically relevant scales.  However,
the shot noise in $P_{gg}$ may be comparable to the galaxy clustering
signal, and so including or excluding the shot noise in $P_{gg}$ makes a
significant difference.

The galaxy-matter cross-correlation coefficient $R$ has been discussed
\citep{dl,p,teg,soda,sw} as a measure of the stochasticity, or scatter, in
the relationship between the galaxy overdensity $\delta_g$ and the matter
overdensity $\delta_m$.  This interpretation of $R$ is clearest when the
Fourier-space representation of $R$ is used, and when the shot noise is
included in $P_{gg}$; thus, we advocate defining $R$ in this manner.  Then
$R(k)$ has a straightforward physical meaning: it is unity on large scales
where $\delta_g(k)$ and $\delta_m(k)$ are simply related, and decreases on
small, nonlinear scales where significant scatter exists in the
relationship.

\section{The standard halo model}\label{halomodel}
Now we will outline the standard halo model, drawing primarily on
\citet{cs} and \citet{s00}.  For the matter power spectrum
$P_{mm}(k)$, the following ingredients are necessary: the linear power
spectrum $P_{\rm lin}(k)$, the number density of haloes as a function
of mass $n(M)dM$, the large-scale bias as a function of halo mass,
$b(M)$, and the average Fourier-transformed density profile of a halo
of mass $M$, $y(k,M)$.  This density profile is normalized to be unity
at $k = 0$:
\begin{equation}
  y(\bk,M) = \frac{1}{M}\int
  \rho(\br,M)\exp(-i\bk\cdot\br)d^3r,
  \label{y_i}
\end{equation}
where $\rho(\br,M)$ is the average real-space density profile of a
halo of mass $M$.  For a spherically-symmetric density profile $\rho(r,M)$,
this becomes
\begin{equation}
  y(k,M) = \frac{4\pi}{M}\int_0^{\infty} \frac{\sin kr}{kr} \rho (r,M) r^2 dr.
  \label{y_sphsym}
\end{equation}
The power spectrum is a sum of one-halo ($1h$) and two-halo ($2h$) terms:
\begin{equation}
  P_{mm}(k) = P_{mm}^{1h}(k) + P_{mm}^{2h}(k),
\end{equation}
where
\begin{equation}
  P_{mm}^{1h}(k) = \int y(k,M)^2 \frac{n(M) M^2}{{\rho}^2} dM,
\end{equation}
and
\begin{equation}
  P_{mm}^{2h}(k) = P_{\rm lin}(k)\left[\int b(M)y(k,M)\frac{n(M)M}{\rho}dM\right]^2.
\end{equation}
Here, $\rho$ is the average matter density.

With galaxies come a few more ingredients: a halo occupation
distribution $(N_{\rm gal}|M)$ giving the distribution of the number of
galaxies inside a halo of mass $M$, giving a total galaxy number
density $n_{\rm gal}$, and a quantity $y_{\rm gal}(k)$, which describes the
average galaxy density profile of a halo, in general different from
its matter density profile.  The galaxy-matter power spectrum $P_{gm}$
and the galaxy power spectrum $P_{gg}$ are also sums of $1h$ and $2h$
terms:
\begin{eqnarray}
  P_{gm}^{1h}(k) & = & \int y(k,M) y_{\rm gal}(k,M)^{p-1}\nonumber\\
  & &\times\frac{\langle N_{\rm gal}|M\rangle}{n_{\rm gal}}\frac{n(M) M}{\rho} dM;\\
  P_{gm}^{2h}(k) & = &
  P_{\rm lin}(k)\int y_{\rm gal}(k,M)\frac{\langle N_{\rm gal}|M\rangle}{n_{\rm gal}} b(M) n(M) dM\nonumber\\
  & &\times\int y(k,M)b(M) \frac{n(M) M}{\rho} dM;\\
  P_{gg}^{1h}(k) & = & \int y(k,M) y_{\rm gal}(k,M)^p\nonumber\\
  & &\times\frac{\langle N_{\rm gal}(N_{\rm gal}-1)|M\rangle}{n_{\rm gal}^2} n(M) M dM;\\
  P_{gg}^{2h}(k) & = & P_{\rm lin}(k)\nonumber\\
  & &\times\left[\int y_{\rm gal}(k,M)\frac{\langle N_{\rm gal}|M\rangle}{n_{\rm gal}}
    b(M) n(M) dM\right]^2.
\end{eqnarray}
Here, $p$ is 2 if there is more than one galaxy per halo, and 1 otherwise;
this arises from an assumption that there is a galaxy at the centre of each
halo of sufficient mass.  (See \citealt{cs} for more explanation.)

\section{The galaxy-halo model}\label{galhalomodel}
A `galaxy halo' is defined in this paper to be a dark-matter halo around a
galaxy.  As in the standard halo model, all power spectra in the
galaxy-halo model are sums of one-halo ($1h$) and two-halo ($2h$) terms.
These could be called one-galaxy and two-galaxy terms, since the
fundamental objects in the galaxy-halo model are galaxies.  However,
galaxies in the galaxy-halo model are pointlike objects.  It is in their
haloes that matter is found, and where the galaxy-matter and matter power
spectra measure the clustering.

Even though the $1h$ and $2h$ terms in the galaxy-halo model share labels
with their counterparts in the standard halo model, the terms in the two
models differ conceptually, and generally differ numerically as well.  That
is, the $1h$ and $2h$ terms get different shares of the total power spectra
in the galaxy-halo model than in the standard halo model.

Since the galaxy power spectrum $P_{gg}$ is one of the inputs of the
galaxy-halo model, the expressions for it are extremely simple.  The
two-halo term $P_{gg}^{2h}$ is just the galaxy power spectrum without shot
noise.  To emphasize that it is a fundamental input into the model, we
simply denote $P_{gg}^{2h}$ as $P$.  There is also a one-halo term in the
full $P_{gg}$, which is the shot noise $1/n$, where $n$ is the total number
density of galaxies.  So, the two terms in $P_{gg}$, with the shot noise
included, are
\begin{equation}
  P_{gg}(k) = P_{gg}^{2h}(k) + P_{gg}^{1h}(k) = P(k) + 1/n.
  \label{pgg}
\end{equation}

Dark-matter halo density profiles affect the other power spectra, $P_{gm}$
and $P_{mm}$.  First we discuss what they are in a simple, single-mass
model, and then in a more realistic model with a distribution of halo
masses.

\subsection{Single-mass model}
In this section appears a highly simplified model, some of the results of
which carry over to the next, more realistic model.  Consider a population
of galaxies of number density $n$ and power spectrum $P(k)$, and with
haloes of identical masses and identical, spherically symmetric density
profiles.  The density profile $\rho(r)$ Fourier-transforms into $y(k)$, in
the same manner as in eq.\ (\ref{y_sphsym}).  The density profile must be
well-behaved in the sense that as $r\to\infty$, $\rho(r)\to0$ in such
a way that $y(k)\to 1$.

Both the $1h$ and $2h$ terms of the galaxy-matter power spectrum depend on
the Fourier-transformed density profile $y(k)$.  The pairs comprising the
$1h$ term are galaxies with dark-matter particles in their own haloes:
\begin{equation} P_{gm}^{1h}(k) = n^{-1}y(k).
  \label{pgm1h_1}
\end{equation}
The $2h$ term is the power spectrum of galaxies with dark-matter particles
in other haloes; it is a product in Fourier space of the galaxy power
spectrum $P(k)$ with the average halo profile $y(k)$.
\begin{equation}
  P_{gm}^{2h}(k) = P(k)y(k).
  \label{pgm2h_1}
\end{equation}
The matter power spectrum is similarly defined:
\begin{equation}
  P_{mm}^{1h}(k) = n^{-1}y(k)^2,
  \label{pmm1h_1}
\end{equation}
\begin{equation}  
  P_{mm}^{2h}(k) = P(k)y(k)^2.
  \label{pmm2h_1}
\end{equation}
This simple model yields simple formulae for the bias $b(k)$ and the
cross-correlation coefficient $R(k)$.  The (squared) bias, excluding the
shot noise from $P_{gg}$, is
\begin{equation}
  b^2_{\rm -sn}(k) = \frac{P(k)}{P_{mm}^{2h}(k) + P_{mm}^{1h}(k)} =
  \frac{P(k)}{P(k) + n^{-1}} y(k)^{-2}.
  \label{b_1}
\end{equation}
If the galaxy power spectrum includes the shot noise, the bias is simply
\begin{equation}  
  b^2(k) = y(k)^{-2}.
  \label{b_1sn}
\end{equation}
The equation for $R^2(k)$ (where $P_{gg}$ includes the shot noise) is even
simpler:
\begin{equation}
  R^2(k) = \frac{[P_{gm}^{2h}(k) +
  P_{gm}^{1h}(k)]^2}{[P(k)+n^{-1}][P_{mm}^{2h}(k) + P_{gm}^{1h}(k)]} = 1.
  \label{R_1}
\end{equation}
This makes sense: galaxies and matter are perfectly cross-correlated if all
galaxy haloes are identical.

\subsection{Multiple-mass model}
Now, more realistically, suppose that there is a distribution of halo
masses and shapes (which need not be spherically symmetric).  As a
pedagogical aid to those familiar with the halo model, we will point out
how various terms change or disappear using the galaxy-halo model.  In
doing this, we do not mean to imply that the galaxy-halo model is a subset
of the halo model, in which additional assumptions are made.  The
fundamental assumptions of the two models differ.

In the halo model, galaxies are put into haloes, while in the galaxy-halo
model, haloes are put around galaxies.  So, in the galaxy-halo model, the
HOD (which does not explicitly appear) is identically 1, and $y_{\rm gal}$
is unnecessary.  The galaxy power spectrum $P$ is the fundamental quantity,
so $P_{\rm lin}$ does not appear.  The bias as a function of $M$, $b(M)$,
also is not needed, since it is subsumed into $P$.  However, there is still
a large-scale bias $b_0$ multiplying the three $2h$ terms; as $k\to 0$,
$P_{gg}^{2h}(k) = b_0P_{gm}^{2h}(k) = b_0^2P_{mm}^{2h}(k)$.

\yeq

Let the total number density of galaxies be denoted $n=\int n(M)dM$, and
the total mass density be denoted $\rho = \int n(M)MdM$.  Here, $n(M)dM$ is
the number density of galaxy haloes of a given mass; henceforth, $n$
denotes the total galaxy number density unless it explicitly appears as a
function of mass.  Not surprisingly, the expressions for $P_{gm}$ and
$P_{mm}$ become more complicated when there is a distribution of masses.
It is still straightforward, though, to express the $1h$ terms.  The
average Fourier-transformed density profile as a function of mass $y(k,M)$
is the average of $y(k)$ over haloes of mass $M$, and the mean-square
$y^2(k,M)$ is the average of $y(k)^2$.  Assuming a large enough volume that
departures from isotropy vanish, these average halo profiles must be
spherically symmetric, and thus have zero imaginary components.  The
density profiles comprising $P_{gm}^{1h}$ are averaged weighting by the
product of the number density of galaxies and the mass density:
\begin{equation} 
  P_{gm}^{1h}(k) = \frac{1}{n\rho}\int{y(k,M)n(M)MdM} =
n^{-1}y_{gm}^{1h}(k),
  \label{pgm1h}
\end{equation}
where $y_{gm}^{1h}$ is a mass-weighted average halo profile
\begin{equation}
  y_{gm}^{1h}(k) \equiv \frac{\int{y(k,M)n(M)MdM}}{\int{n(M)MdM}}.
  \label{ygm1h}
\end{equation}
The density profiles comprising $P_{mm}^{1h}$ are averaged weighting
by the mass squared:
\begin{equation}
  P_{mm}^{1h}(k) = \frac{1}{\rho^2}\int{y^2(k,M)n(M)M^2dM} =
  \frac{\mu}{n} y_{mm}^{1h}(k)^2,
  \label{pmm1h}
\end{equation}
where 
\begin{equation}
  y_{mm}^{1h} \equiv \sqrt{\frac{\int{y^2(k,M)n(M)M^2dM}}{\int{n(M)M^2dM}}},
  \label{ymm1h}
\end{equation}
and $\mu$ is a dimensionless mean-square halo mass,
\begin{equation}
  \mu\equiv \langle M^2\rangle/\langle M\rangle^2 \ge 1.
  \label{mu}
\end{equation}

Not as much can be said about the $2h$ terms with the assumptions made so
far, primarily because the large-scale bias factor of a set of haloes
varies with their mass.  This behaviour can be modeled \citep[e.g.][]{mw},
but to our knowledge, such a model has not been formulated which counts
subhaloes as haloes, as in the galaxy-halo model.  Even if there were such
a model, adding it to the galaxy-halo model would cause a significant
increase in the galaxy-halo model's complexity, which we wish to avoid.

However, one thing may be safely assumed about the $2h$ terms: on
large-enough scales (where the $1h$ terms are negligible, and all average
Fourier-transformed density profiles are unity), the cross-correlation
coefficient $R(k)=1$.  To see this, suppose the universe consists of
regions large enough so that the galaxy bias in each region is entirely
local \citep{coles} and is statistically independent of the bias in other
regions.  In the limit $k\to 0$, a measurement of $R(k)$ requires averaging
over a number of regions which goes to infinity, squeezing the variance of
the biases in different regions to zero.  This gives $R(k\to 0) = 1$.
Assuming as much, the following equations hold:
\begin{eqnarray}
  P_{gm}^{2h}(k) &=& P(k)\frac{y_{gm}^{2h}(k)}{b_0}\label{pgm2h};\\
  P_{mm}^{2h}(k) &=& P(k)\left[\frac{y_{mm}^{2h}(k)}{b_0}\right]^2\label{pmm2h}.
\end{eqnarray}
Here, $b_0 \equiv \lim_{k\to 0}\sqrt{P(k)/P^{2h}_{mm}(k)}$ is a large-scale
bias (usually of order unity) and $y_{gm}^{2h}(k)$ and $y_{mm}^{2h}(k)$ are
effective average halo density profiles, defined to be unity as $k\to 0$.
These average halo density profiles have no imaginary component, by
construction, since they are defined in eqs.\ (\ref{pgm2h}) and
(\ref{pmm2h}) in terms of other real quantities.

In general, $y^{1h}$ differs from $y^{2h}$ (for both galaxy-matter and
matter power spectra) because halo density profiles change systematically
with the clustering strength of the haloes, and $y_{gm}$ differs from
$y_{mm}$ (for both $1h$ and $2h$ terms) because the $y_{mm}$ is a
root-mean-square average, whereas $y_{gm}$ is a straight average.  Even
though these terms typically differ, below we will explore what emerges
under the approximations that $y^{1h}=y^{2h}$ and $y_{gm}=y_{mm}$.

Figure \ref{yeq} compares these four average Fourier-transformed density
profiles $y_{gm,mm}^{1h,2h}$, for galaxies placed at the centres of two
different sets of haloes from $N$-body simulations, characterized by large
(top) and small (bottom) mass thresholds.  For each panel, the profiles do
not correspond exactly, but they are similar.  The agreement is better
using a large mass threshold because excluding small haloes narrows the 
distributions of halo masses and density profiles.

Assuming that $y_{gm}^{1h}=y_{gm}^{2h}\equiv y_{gm}$, and
$y_{mm}^{1h}=y_{mm}^{2h}\equiv y_{mm}$, the expressions for $P_{gm}$
and $P_{mm}$ simplify:
\begin{eqnarray}
  P_{gm}(k) &\approx& [P(k)/b_0 + 1/n]y_{gm}(k);\label{pgm}\\
  P_{mm}(k) &\approx& [P(k)/b_0^2 + \mu/n]y_{mm}(k)^2\label{pmm}.
\end{eqnarray}

With these power spectra in hand, it is possible to calculate ratios
between them: the bias $b(k)$, the cross-bias $b/R = P_{gg}/P_{gm}$, and
the cross-correlation coefficient $R(k)$.  Here are equations for the bias
(squared) $b^2 \equiv P_{gg}/P_{mm}$, excluding the shot noise from
$P_{gg}$:
\begin{eqnarray}  
  b^2_{\rm -sn}(k) & = & \frac{P}{P(y_{mm}^{2h}/b_0)^2 + (\mu/n)(y_{mm}^{1h})^2}\\
  & \approx & \frac{P}{P/b_0^2+\mu/n}{y_{mm}}^{-2}.
  \label{b}
\end{eqnarray}
Equation (\ref{b}) follows if $y_{mm}^{1h}=y_{mm}^{2h}$.  If the shot noise
is included in $P_{gg}$, then eq.\ (\ref{b}) becomes
\begin{equation}
  b^2(k) \approx \frac{P+n^{-1}}{P/b_0^2+\mu/n}{y_{mm}}^{-2}.
  \label{bpsn}
\end{equation}

Equation (\ref{b}) contains the general features in the bias (excluding the
shot noise from $P_{gg}$) which have been observed from simulations and
from the halo model (Seljak 2000).  On large scales, the halo profiles in
real space go to zero [and thus $y_{mm}(k)\to 1$], and the galaxy-matter
two-halo term overwhelms the one-halo term [$P(k)/b_0^2 \gg \mu/n$].  Thus,
as expected, the only signal on large scales is the large-scale bias, and
$b_{\rm -sn}(k)=b_0$.  On intermediate scales, where the two-halo and
one-halo terms in the matter power spectrum are comparable [$P(k)/b_0^2
\approx \mu/n$], assuming that $y_{mm}(k) \approx 1$ still holds, the bias
decreases.  On small scales, the one-halo term dominates [$P(k)/b_0^2 \ll
\mu/n$], and so the behaviour of the bias depends on whether $P(k)$ or
$y_{mm}(k)^2$ decreases faster as $k\to\infty$.  Generally, $y_{mm}(k)^2$
decreases faster, forcing the bias upward.  If $P_{gg}$ includes the shot
noise, then the bias always increases with $k$ at small scales.

The cross-bias $b/R$, keeping the shot noise in $P_{gg}$, is
\begin{eqnarray}
  \frac{b}{R} & \equiv & \frac{P_{gg}}{P_{gm}} = \frac{P+n^{-1}}{Py_{gm}^{2h}/b_0 + n^{-1}y_{gm}^{1h}}\\
  & \approx & \frac{P + n^{-1}}{P/b_0 + n^{-1}}{y_{gm}}^{-1}.
  \label{bore}
\end{eqnarray}
Equation (\ref{bore}) follows if $y_{gm}^{1h}=y_{gm}^{2h}$.

A formula with few inputs also emerges for the cross-correlation
coefficient $R^2 \equiv P_{gm}^2/(P_{gg} P_{mm})$.  Including the shot
noise in $P_{gg}$,
\begin{eqnarray}
R^2(k) & = &\frac{(Py_{gm}^{2h}/b_0 + n^{-1}y_{gm}^{1h})^2}
  {(P+n^{-1})[P(y_{mm}^{2h}/b_0)^2+(\mu/n)(y_{mm}^{1h})^2]}\\
  \label{r2ay}
  & \approx &\frac{(P/b_0 + n^{-1})^2}{(P+n^{-1})(P/b_0^2+\mu/n)}
  \left(\frac{y_{gm}}{y_{mm}}\right)^2\\
  & \approx &\frac{(P/b_0 + n^{-1})^2}{(P+n^{-1})(P/b_0^2+\mu/n)}.
  \label{r2a}
\end{eqnarray}
Equation (\ref{r2ay}) is true if $y_{mm}^{1h}=y_{mm}^{2h}$ (and similarly
for $y_{gm}$), and for eq.\ (\ref{r2a}), a more extreme assumption is made,
that $y_{gm} = y_{mm}$.  Since $y_{gm}$ is weighted by mass, and $y_{mm}$
by mass squared, the latter assumption will only be valid if halo profiles
do not depend on mass, which is almost certainly not the case.  Excluding
the shot noise in $P_{gg}$ turns $(P+n^{-1})$ into $P$ in the denominator
in eqs.\ (\ref{r2ay}) and (\ref{r2a}).

In eq.\ (\ref{r2a}), on large scales, $P$ dominates both $n^{-1}$ and
$(\mu/n)$, so $R^2$ approaches unity.  On small scales, $P$ is small, and
so $R^2$ approaches $1/\mu$.  More specifically, under these
approximations, $R^2$ is an interpolation between 1 (its $2h$ value) and
$1/\mu$ (its $1h$ value), weighted by the respective matter terms:

\begin{eqnarray}
  R^2(k) & \approx &
  \frac{\left[\left(P+\frac{1}{n}\right)\left(\frac{P}{b_0^2}+\frac{1}{n}\right) -
  \frac{P}{n}\left(1-\frac{1}{b_0}\right)^2\right]{y_{gm}}^2}{(P+n^{-1})(P_{mm}^{2h}+P_{mm}^{1h})}\\
  & \approx & \frac{P_{mm}^{2h} + (1/\mu)P_{mm}^{1h}}{P_{mm}^{2h}+P_{mm}^{1h}} - \epsilon(b_0),
  \label{r2m}
\end{eqnarray}
where we assume that $y_{gm}^{1h}=y_{mm}^{1h}$ (and similarly for the $2h$ terms).  Here, $\epsilon(b_0)$ is small if
$b_0 \approx 1$:
\begin{equation}
  \epsilon(b_0) = \frac{(b_0 - 1)^2}{1+nP + \mu b_0^2[1+1/(nP)]}.
\end{equation}

Although eq.\ (\ref{r2a}) is exactly true in the more general case of zero
variance in halo density profile shape, it may be helpful to interpret eq.\
(\ref{r2a}) by visualizing the haloes as a collection of nuggets (instead
of extended haloes) of varying mass.  Consider the smallest scales, $k
\ga\pi/r_{\rm min}$, where $r_{\rm min}$ is the smallest intergalactic
distance.  Here, the Fourier-transformed galaxy and matter overdensities
$\delta_g(k)$ and $\delta_m(k)$ sample at most one galaxy, and so $R^2(k)$
sees only the mass dispersion of galaxies.  This is where the galaxy power
spectrum $P\ll 1/n$, and thus $R^2\to 1/\mu$ in eq.\ (\ref{r2a}).  At
progressively larger scales, $R^2$ samples more and more galaxies, until at
the largest scales, so many galaxies enter the average that the scatter in
the $\delta_g$-to-$\delta_m$ relationship vanishes, pushing $R^2\to 1$.

\subsection{Orphans}\label{orph}

Up to now, we have only defined a galaxy halo as a clump of dark matter
(surrounding a galaxy) whose density falls off to zero at large radius.  In
practice, one way to define a galaxy halo sample could be to populate a set
of bound dark-matter haloes and subhaloes (from a simulation, for example)
with galaxies.  This set of galaxy haloes could be characterized by a
bound-mass or circular-velocity threshold, for example.  But what about
orphaned matter particles which are not bound to any galaxy halo?  Such
matter exists not only in voids, but in small, isolated haloes which do not
meet the criteria to host a galaxy and are not bound to any larger galaxy
haloes.

One way to deal with orphans is to adopt them into galaxy haloes, removing
the condition that galaxy haloes must be gravitationally bound.  However,
it is not clear how to partition the unbound matter into galaxies, and
altering the partition could significantly affect the dimensionless
mean-square halo mass $\mu$.

In this paper, we exclude orphans from galaxy haloes, but they cannot be
completely ignored.  Even under the assumption that orphans do not affect
clustering properties, excluding them in the calculation of the matter and
galaxy-matter power spectra results in the wrong normalization.  If
$P_{gm}$ and $P_{mm}$ are calculated using only non-orphans, then the
multiplicative `orphan factor' $\rho_h/\rho$ must be applied once to
$P_{gm}$ and twice to $P_{mm}$ to get the normalization right.  Here,
$\rho_h$ is the density of matter in galaxy haloes, and $\rho$ is the total
matter density.  `Orphan factors' must be used with the bias or cross-bias,
but they cancel out for the cross-correlation coefficient $R$.

Even if orphans are excluded, the question of partitioning the matter into
galaxy haloes can be ambiguous.  Power spectra care only about density
contrasts, not about whether matter is bound to galaxies, so boundedness is
not necessarily the right test to determine galaxy halo membership.

In the context of the galaxy-halo model, the best partition of matter is
the one which makes the predictions of the galaxy-halo model work best.
There are a few ways of judging matter partitions using this criterion.
One of them is to make the approximations $y_{gm}^{1h} = y_{gm}^{2h}$ and
$y_{mm}^{1h} = y_{mm}^{2h}$ hold as closely as possible, but this is hard
to test over a wide range of halo samples.  The easiest meaningful quantity
to measure from a partition is the dimensionless mean-square mass $\mu$.
Measuring $R^2(k)$, which is independent of the partition, and looking at
its typical small-scale value, gives an idea of the `natural' dimensionless
mean-square halo mass.  However, as displayed below in Figure \ref{scdr},
if haloes extend out to a scale where $P_{gg}$ is significant, $R^2$ may
stay above its characteristic small-scale value, making $\mu$ hard to
determine from $R^2$.  There is some ambiguity in how precisely the matter
should be partitioned, but that is not necessarily a bad thing, since the
observed power spectra cannot depend on the partition.

\section{Tests}
Can the galaxy-halo model be used to extract meaningful information from
observations?  The galaxy-halo model contains three items which are
potentially useful: simple descriptions of the galaxy-matter bias and
cross-correlation coefficient; and the capacity to separate out $1h$ and
$2h$ terms from an observed galaxy-matter power spectrum.  This section
describes tests of these items, and also explores how predictions of the
galaxy-halo model vary with properties of a galaxy-halo population.

\subsection{Mock halo catalogs}\label{mhc}
\subsubsection{Isolating $\xi_{gm}^{1h}$}

\nfwhp

The galaxy-matter correlation function $\xi_{gm}$ is sometimes interpreted as
a measure of the average overdensity profile $\bar\delta(r)$ of haloes
around galaxies.  However, the one-halo $\xi_{gm}^{1h}$ may be a more
appropriate measure of the average overdensity profile of galaxy haloes,
since including $\xi_{gm}^{2h}$ would double-count matter in overlapping
regions.  The $1h$ and $2h$ terms of $\xi_{gm}$ are not observable by
themselves; only their sum is.  The galaxy-halo model for $P_{gm}$ allows
removal of an effective two-halo contribution to the galaxy-matter
correlation function $\xi_{gm}$, if the galaxy correlation function
$\xi_{gg}$ is known as well.

In this section, we discuss a test of how well the average overdensity
profile $\bar\delta(r)$ can be measured from $\xi_{gg}$ and $\xi_{gm}$
within the framework of the galaxy-halo model.  For galaxy positions, we
used the centres of dark-matter haloes from a $256\hMpc$, 256$^3$-particle
{\LCDM} dark-matter-only $N$-body simulation \citep[][hereafter NHG]{nhg}.
To detect the haloes, we used the halo-finding algorithm {\VOBOZ}
\citep{voboz}, with a density threshold of 100 times the mean density.  All
haloes exceeded a $2\sigma$ {\VOBOZ} probability threshold.  The closest
pair of galaxies was separated by $0.7\hMpc$.

Around the galaxies, we put identically shaped NFW \citep{nfw97} profiles
with scale radii of $1\hMpc$, all truncated at a deliberately large radius
of $8\hMpc$.  At this truncation radius, many haloes overlapped, providing
a sizeable $2h$ term to subtract off from $\xi_{gm}$.  Although all the
haloes had identical shapes, we preserved the number of particles in each
halo from the simulation by varying the density profiles in the mock
catalog by multiplicative constants.  The haloes ranged in particle number
from 821 to 10222 particles, with a dimensionless mean-square mass $\mu =
1.39$, for a total of 917501 particles.

\scdr

We used the following procedure to separate the $1h$ and $2h$ terms of
$\xi_{gm}$.  First, Fourier-transform $\xi_{gm}$ and $\xi_{gg}$ [e.g.\
using {\FFTLog} \citep{h2000}], and then solve for $y_{gm}$ in eq.\
(\ref{pgm}).  To obtain $\xi_{gm}^{1h}$, Fourier-transform $P_{gm}^{1h} =
y_{gm}^{1h}/n$ back into real space.  Doing this requires an estimate of
the large-scale bias $b_0$.  If the size of the haloes is small compared to
the volume of the sample, it is safe to assume that $y_{gm}(k_{\rm min})$
is of order unity, where $k_{\rm min}$ is the smallest wavenumber in the
power spectrum.  An upper limit, and quick estimate, of $b_0$ comes from
assuming that $y_{gm}(k_{\rm min})=1$.  If both $\xi_{gg}$ and $\xi_{gm}$
are measured well out to linear scales, $b_0$ may be measured directly from
their ratio.

In Fig.\ \ref{nfwhp}, the full $1+\xi_{gm}$ overestimates the NFW profile
by almost a factor of 2 at the largest scales, whereas $1+\xi_{gm}^{1h}$
reproduces it much better.  The full and $1h$ $\xi_{gm}$'s start to diverge
at about $r = 0.7\hMpc$, which is the separation of the closest pair of
galaxies, where $1+\xi_{gm}^{2h}$ starts to be positive.  Our initial try
of $y_{gm}(k_{\rm min})=1$ (giving $b_0 = 0.97$) subtracted off most of
$\xi_{gm}^{2h}$, but using $y_{gm}(k_{\rm min})=0.8$ (giving $b_0 = 0.72$)
resulted in a better fit.  Thus, even though the halo diameters were only
$1/16$ of the box size $l$, evidently $y_{gm}(k_{\rm min} = 2\pi/l)$ did not
quite reach unity.

\subsubsection{Behaviour of the cross-correlation coefficient}
This section describes a test of the galaxy-halo model approximations for
the squared cross-correlation coefficient $R^2$, eqs.\ (\ref{r2ay}) and
(\ref{r2a}).  We wanted to gauge the accuracy of the approximation, as well
as investigate how halo profiles affect $R^2$ in general.  For this test,
the galaxy-halo profiles for all masses had a fixed density at each radius,
but the truncation radius varied with mass.  We used a convenient density
profile, $\rho(r)\sim r^{-2}$, for which the truncation radius is
proportional to mass.

For this test, the galaxy positions were those of the centres of the 4132
largest {\VOBOZ} haloes exceeding $2\sigma$ in the $256\hMpc$ simulation
described above.  The haloes ranged in mass from 230 to 10222, enough to
give a dimensionless mean-square mass $\mu=2$.  The smallest galaxy
separation was $0.113\hMpc$.

Figure \ref{scdr} shows the results of varying the halo radius per particle
from $10^{-5.5}$ to $10^{-2.5}~\hMpc$.  The power spectra $P_{gg}$,
$P_{gm}$ and $P_{mm}$ were calculated from 3D FFTs of the galaxy and matter
distributions, reaching small scales by `folding' the particle distribution
by factors of two (Klypin, private communication).  For each fold, the
boxes were split into eight octants, and each octant was superposed
together in a box of half the size; thus, each fold enabled the FFT to
reach scales smaller by a factor of two.  At large scales, $R^2\approx 1$
because with a large window function, many galaxy haloes are averaged over.
The $R^2$ curves then descend with $k$ as eq.\ (\ref{r2a}) predicts, but
then turn up at about the scale ($\pi/r$) of the largest halo radius, and
finish ascending at about the scale of the smallest halo radius.  It makes
sense that $R^2\approx 1$ at small scales where the haloes are identical.
Knowing everything about every halo makes it possible to calculate
$y_{gm}/y_{mm}$ analytically; putting this into eq.\ (\ref{r2ay}) brought
the approximation quite close to the measured $R^2$.  For some reason, the
measured $R^2$ curves did not quite reach unity, as the analytical curves
would predict.  Downturns, only visible here for the two largest halo radii
per particle, occur at about the scale of the tightest matter pair.  Such
downturns would not occur in the real Universe, which has much higher
`resolution.'

\simr

\subsection{Simulations}\label{sims}

The following tests involve more realistic density fields, drawn from
simulations.  The tests evaluate the galaxy-halo model's descriptions of
the cross-correlation coefficient and the bias between galaxies and matter.

\subsubsection{The cross-correlation coefficient approximation}
In this section, we discuss a comparison of the predictions of
approximation (\ref{r2a}) for the squared cross-correlation coefficient
$R^2(k)$ with measurements from simulations.  The test also investigates
the degree to which it matters if orphans (particles not gravitationally
bound to haloes) are included in the calculation.  The galaxy positions
were drawn from the centres of {\VOBOZ}-identified haloes in a suite of
four {\LCDM} simulations described by NHG.

Figure \ref{simr} shows $R^2$ for various halo catalogs, using different
box sizes (32, 64, 128, and 256 $\hMpc$), and different lower mass
thresholds.  An additional threshold of $2\sigma$ in {\VOBOZ} halo
probability eliminated many spurious haloes.  Measurements of $R^2$, both
including and excluding orphans (see sect.\ \ref{orph}), are shown.  Orphans
make a significant difference in $R^2$ with a high halo mass threshold, but
not otherwise.  Orphans would likely make a greater difference if there
were an upper mass threshold as well, since most of the pairs comprising
$P_{gm}$ and $P_{mm}$ lie in the largest haloes.

Figure \ref{simr} also shows the approximation of eq.\ (\ref{r2a}).  For
the dashed curves, we measured the dimensionless mean-square halo mass
$\mu$ using the {\VOBOZ} particle halo membership.  Some particles belong
to more than one halo in {\VOBOZ}; we removed this ambiguity by assigning
each particle to the smallest-mass {\VOBOZ} halo containing it.

In Fig.\ \ref{simr}, especially for low mass thresholds, the $R^2$ curves
do not reach their characteristic small-scale value as predicted from the
{\VOBOZ} $\mu$.  The dashed curves use a reduced $\mu^\prime$, in which the
masses of haloes in clusters are equalized in an extreme way.  (We define a
cluster to be a set of haloes such that each halo is within the half-mass
radius of another halo in the cluster.)  For $\mu^\prime$, the mass of each
halo in a cluster is set to the mass of the parent halo (the largest halo
in the cluster) divided by the number of haloes in the cluster.  This is
not an unreasonable thing to do since, from the point of view of $R^2$, it
may not be appropriate to distinguish between large parent haloes and small
subhaloes.  In a cluster environment, all $R^2$ sees is a group of galaxies
surrounded by a bunch of matter; it does not know whether the matter is
bound to parent haloes, to subhaloes, or to neither.

Figure \ref{difmu} shows $\mu$ and $\mu^\prime$ as a function of lower halo
mass cut-offs.  As expected, the difference between them grows with the
amount of substructure, i.e.\ as the halo mass cut-off is decreased.
Figure \ref{difmu} also shows $\mu_{\rm meas}$ as measured from the curves
in Fig.\ \ref{simr}; $\mu_{\rm meas} = 1/R^2_{\rm min}$, where $R^2_{\rm
min}$ is the lowest value $R^2(k)$ attains for $k<10 \ihMpc$.  A simple fit
to $\mu_{\rm meas}$ from our simulations, shown as the dotted curve in
Fig.\ \ref{difmu}, is
\begin{equation}
  \mu_{\rm meas} \approx 1+\left(\frac{M_{\rm min}}{10^{13}\hMsun}\right)^{-0.4},
  \label{mumeas}
\end{equation}
where $M_{\rm min}$ is the lower halo mass cut-off of the sample.

\difmu
\vspace{1 cm} 
The modified (in an extreme fashion) $\mu^\prime$ does agree with $\mu_{\rm
meas}$ better than the original $\mu$, but there is still a significant
difference, which can be explained with reference to Fig.\ \ref{scdr}.
Galaxy-halo profiles encroach into the regime where $P_{gg}$ is
significant, causing an upturn before (i.e.\ at larger scales than) $R^2$
would have attained its smallest value.  Another way to look at the generic
rise of $R^2$ at small scales is that, again referring to Fig.\ \ref{scdr},
the profiles of galaxy haloes of different masses are more similar to each
other at small scales than at intermediate scales.

Figure \ref{difmu} also shows an estimate of $\mu$ from a \citet{st99} mass
function with a lower, but not upper, mass limit.  This estimate agrees
with $\mu_{\rm meas}$ about as well as the {\VOBOZ} $\mu$ estimate, except
at high masses.  We also tried estimating $\mu$ from the \citet{st99} mass
function using an upper mass limit given by the largest halo mass appearing
in each simulation.  These results are not shown; this procedure estimated
$\mu_{\rm meas}$ well for high-mass (and therefore low-substructure) halo
samples in the $256 \hMpc$ simulation, but somewhat poorly for lower-mass
haloes in other simulations.

\subsubsection{The bias approximation}

\simb

Figure \ref{simb} shows the galaxy-matter bias $b_{\rm -sn}^2 =
P_{gg}/P_{mm}$ (excluding the shot noise from $P_{gg}$) as measured from a
set of haloes taken from a $32\hMpc$ simulation (NHG), with a physical mass
cut-off of $3\times 10^{10}\hMsun$ (128 particles).  On small scales, the
measured bias increases relative to what the galaxy-halo model would
predict if the haloes were infinitesimal nuggets of mass [i.e.\ that
$y_{mm}(k)=1$ in eq.\ (\ref{b})].  This occurs because haloes are extended
objects, i.e.\ because ${y_{mm}}^2$, the mean-square (weighting by the halo
mass squared) halo profile, decreases from unity at small scales.

For this figure, we used the {\VOBOZ} estimate of the dimensionless
mean-square halo mass $\mu=45$ (see the bottom middle panel of Fig.\
\ref{simr}).  The approximate concordance of the two quantities in the top
panel reflects the fact that $y_{mm}$ is, as expected, near unity on large
scales.  To get the normalization of $P_{mm}$ right, it was necessary to
divide by `orphan factors' (see section \ref{orph}) $(\rho_h/\rho)^2$,
where $\rho$ is the total matter density, and $\rho_h$ is the density of
matter in galaxy haloes.  We fitted $b_0 = 0.27$ by requiring that the
result of eq.\ (\ref{b}), after dividing by $(\rho_h/\rho)^2$, equal the
measured bias curve in the largest-scale bin.  Including orphan factors,
the effective large-scale bias becomes 0.75, rather small because the mass
cut-off is low.

The bottom panel shows ${y_{mm}}^2$, the quotient between the measured bias
and the infinitesimal-nugget prediction in eq.\ (\ref{b}), along with
measurements of $(y_{mm}^{1h})^2$ and $(y_{mm}^{2h})^2$.  These would all
lie on top of each other if the assumption that $y_{mm}^{1h} = y_{mm}^{2h}$
used for eq.\ (\ref{b}) were true.  For this set of haloes, $y_{mm}$ does
not particularly trace the empirical $y_{mm}^{2h}$ [eq.\ (\ref{pmm2h})],
but, conveniently, it does seem to follow the better-defined $y_{mm}^{1h}$
[eq.\ (\ref{ymm1h})].  This is not surprising, since the regime where
$y_{mm}$ is interesting (i.e.\ not unity) is on small scales, where
$P_{mm}^{1h}$ dominates $P_{mm}^{2h}$.  At least in this case, the
galaxy-halo model explanation of the bias rising on small scales because of
an effective halo profile works plausibly well.

\section{Interpreting observations}

Increasingly sophisticated observations of weak gravitational lensing have
recently led to high-signal-to-noise measurements of galaxy-matter
clustering \citep{hoek, sheldon}.  Measurements of the galaxy-matter
correlation function $\xi_{gm}$ have previously been interpreted in at
least two fashions: by direct comparison with haloes in dark-matter
simulations such as the ones we have used \citep{iro}, and in the context
of the halo model \citep{gs02, s05a, manbaum}.

In this section, we illustrate how the galaxy-halo model can be used to
extract information from measurements of $\xi_{gm}$ and $\xi_{gg}$;
specifically, we use $\xi_{gm}$ \citep{sheldon} and $\xi_{gg}$ \citep{z02}
as measured from a volume-limited sample of luminous Sloan Digital Sky
Survey (SDSS) galaxies.  Sheldon et al.\ and Zehavi et al.\ have also made
more precise measurements from a larger, flux-limited sample, but they are
harder to interpret, since the luminosity cut-off and galaxy number density
change with redshift.  On small scales where measurements exist for
$\xi_{gm}$ but not for $\xi_{gg}$, we extrapolated $\xi_{gg}$ with a power
law based on the two smallest-scale points.  We also tried assuming
$\xi_{gg}=0$ on small scales, which changed the results negligibly.

There are at least two ways to extract useful information from these
observations with the galaxy-halo model.  First, observations give the
cross-bias $b(k)/R(k)$; with a model of the galaxy-matter cross-correlation
coefficient $R(k)$, it is possible to measure the bias $b(k)$.  Second, it
is possible to extract an effective $\xi_{gm}^{1h}(r)$, or average
overdensity profile $\bar\delta(r)$, of haloes around galaxies, as in
Figure \ref{nfwhp}.

\subsection{Measurement of bias}
To measure the bias from the cross-bias $P_{gg}(k)/P_{gm}(k) = b(k)/R(k)$,
it is necessary to estimate the cross-correlation coefficient $R(k)$.  With
a measurement of the bias, it is then possible to obtain the matter power
spectrum.  To get $R(k)$, it makes sense to use the simplest expression for
$R(k)$ the galaxy-halo model has to offer, eq.\ (\ref{r2a}).  The most
brazen assumption used for this equation is that $y_{gm}(k) = y_{mm}(k)$.
As shown in Fig.\ \ref{scdr}, if halo profiles vary systematically with
mass (which they almost certainly do in the real Universe), this assumption
is valid only on scales larger than that of the largest halo.  In
simulations (Fig.\ \ref{simr}), the approximation is good for $k \la 3
\ihMpc$, which makes sense since clusters have real-space sizes $\la 1\hMpc
\approx \pi/(3\ihMpc)$.  If an accurate model for $y_{gm}/y_{mm}$ is added,
it would allow an accurate estimate of $R^2$ on smaller scales, using eq.\
(\ref{r2ay}).

Eq.\ (\ref{r2a}) allows estimation of $R^2(k)$ with four ingredients: the
galaxy power spectrum $P$, the galaxy number density $n$, a large-scale
bias $b_0$, and the dimensionless mean-square halo mass $\mu$.  The first two
of these are known from the $\xi_{gg}$ measurement, but $\mu$ and $b_0$ are
not.

Figure \ref{difmu} suggests that the best way to estimate $\mu$ is to
measure it from haloes in a simulation.  We detected haloes with {\VOBOZ}
in the same $256\hMpc$ simulation as used for Fig.\ \ref{simr} at redshift
0.1; the redshift of the observed sample varies between $0.1<z<0.174$.  In
a list of the haloes exceeding a $2\sigma$ probability threshold, the most
massive 10028 haloes gave the same number density ($6\times
10^{-4}~h^3\,{\rm Mpc}^{-3}$) as the observed sample.  The haloes ranged in
particle number from 111 to 8364 (physically, from $9\times 10^{12}$ to
$7\times 10^{14}\hMsun$), giving $\mu=2.2$.

The correlation functions $\xi_{gg}$ and $\xi_{gm}$ from this simply
defined set of haloes agree quite well with their observed counterparts.
Previously, \citet{iro} compared the observed correlation functions to
those of sets of haloes in their own simulations.  Using a simple halo mass
cut-off such as ours, their simulated correlation functions were
significantly higher than the observations.  However, by using a reasonable
scatter in the mass-luminosity relation of dark matter haloes, they were
able to lower the theoretical correlation functions to match the
observations more closely.  Such a `fuzzy' halo mass cut-off lowered the
correlation functions by allowing smaller-mass (and more weakly clustered)
haloes into the sample.  We do not fully understand why our correlation
functions were lower (and thus able to reproduce the observations using a
simple mass cut-off), but we suspect it might be explained largely from the
small value of $\sigma_8=0.63$ used in our simulations.

It would be useful to estimate $\mu$ without taking the time to run and
analyse a simulation.  One alternative might be to estimate $\mu$ from a
mass function \citep[e.g.][]{st99}.  However, there is no reason to expect
this to work perfectly, since the haloes in such a mass function are
`largest virialized' haloes as in the standard halo model.  In Fig.\
\ref{difmu}, we show how well one attempt at estimating $\mu$ from this
mass function works; it gives $\mu$ to within a factor of 2 or so.  It may
be possible to improve this guess by fixing an upper halo mass cut-off in
addition to the lower mass cut-off we used.  Another way to improve the
$\mu$ estimate might come from, for example, combining a halo mass
function, a subhalo mass function, and a halo occupation distribution, but
with all of these ingredients, the galaxy-halo model would approach the
complexity of the standard halo model.

What about the large-scale bias $b_0$?  If $P_{gm}$ and $P_{gg}$ are
measured well into the linear regime, where one is confident that $y_{gm} =
1$, then $b_0$ may be measured directly from $P_{gm} = P/b_0 + n^{-1}$.
However, at present $P_{gm}$ is not measured to such large scales, so to
analyze the present observations, it is necessary to make an educated guess
for $b_0$.

Given the dimensionless mean-square halo mass $\mu$, putting $b_0 = 1/\mu$
in eq.\ (\ref{r2a}) gives the largest-possible $R^2(k)$ at each $k$, giving
\begin{equation}
  R^2_{b_0=1/\mu} = \frac{1/\mu + nP}{1+nP}.
  \label{maxr}
\end{equation}
The smallest-possible $R^2(k)$ occurs when $b_0 \to \infty$ on large scales
where $nP>1/\mu$, and when $b_0 \to 0$ on small scales where $nP<1/\mu$.
These minimum values of $R^2$ are
\begin{eqnarray}
  R^2_{b_0\to\infty} & = & [1+1/(nP)]^{-1};\\
  R^2_{b_0\to 0} & = & [\mu (nP+1)]^{-1}.
  \label{minr}
\end{eqnarray}

\iditr
\iditb

Since in the present case, $P_{gm}$ is unavailable on confidently linear
scales, the best way to get $b_0$ seems to be, again, to find it from a
simulation.  The $b_0$ we used comes from fitting eq.\ (\ref{r2a}) to the
actual $R^2(k)$ in the lowest-wavenumber bin, giving $b_0 = 1.12$.  With no
other information, a reasonable zeroth-order guess would be $b_0 = 1$.

Figure \ref{iditr} shows how $R^2(k)$, as calculated with eq.\ (\ref{r2a}),
varies with $b_0$.  It is unlikely that $b_0$ would wander by more than a
factor of two or so from unity.  In the lowest-wavenumber bin, $R$ varies
only by a factor of $\sim 1.2$ as $b_0$ varies between $1/2$ and $2$.  The
measured $R^2$ from the simulation appears as the dashed curve.

Figure \ref{iditb} shows the bias, both including and excluding the shot
noise from $P_{gg}$, as inferred by dividing $R/b$ by $R$ from eq.\
(\ref{r2a}).  We plot $1/b$ instead of $b$ (as Sheldon et al.\ do) because
the error bars are larger in $P_{gm}$ than in $P_{gg}$.  The squares are
the raw observations, $R/b$, and the dashed line shows the result after
dividing this by $R$.  In calculating $R$, $\mu=2.2$ and $b_0=1.12$ are
fixed, but the thick error bars floating in the upper-left corners show the
largest fluctuations (which occur at the largest scales) in $1/b_{\rm -sn}$
and $1/b$ if $b_0$ is multiplied and divided by 2.

The thin, larger error bars are the observational error bars in $1/b$,
propagated through from $\xi_{gm}$ and $\xi_{gg}$.  The error bars on
$P_{gm}$, denoted $\delta P_{gm}$, are obtained by putting the covariance
matrix of $\xi_{gm}$ (which Erin Sheldon kindly provided to us) through a
two-dimensional {\FFTLog}.  Unfortunately, rigorous error bars have not
been measured for $P_{gg}$ for this sample.  As suggested by Idit Zehavi
(private communication), we crudely estimated the error bars on $\xi_{gg}$
by assuming that in each bin, the fractional error
$(\delta\xi_{gg})/\xi_{gg}$ is the same as that in the angular galaxy
correlation function $(\delta w_p)/w_p$, whose error bars have been
measured.  To estimate $\delta P_{gg}$, we put both $\xi_{gg} \pm
\delta\xi_{gg}$ through {\FFTLog}; we set $\delta\xi_{gg} = [F(\xi_{gg} +
\delta\xi_{gg})-F(\xi_{gg} - \delta\xi_{gg})]/2$, where $F$ denotes a
Fourier transform.  The crudeness of $\delta P_{gg}$ is not terribly
worrisome, though, since $(\delta P_{gm})/P_{gm} \gg (\delta
P_{gg})/P_{gg}$.

The solid curves in Fig.\ \ref{iditb} show the bias as measured from the
simulation.  There is good agreement between the dashed and solid curves on
large scales, $k \la 3 \ihMpc$, where eq.\ (\ref{r2a}) works reasonably
well.  It should be kept in mind that comparing the dashed to the solid
curves tests not the galaxy-halo model, but how well the set of haloes
chosen from the simulation represents the observed galaxies.

\subsection{Measurement of average halo density profiles}\label{avgdp}

Now we describe a measurement of an average halo density profile from these
observations of $\xi_{gg}$ and $\xi_{gm}$; for a description of the
procedure, see section \ref{mhc}.

\idithp

Figure \ref{idithp} shows a splitting of the observed galaxy-matter
correlation function $\xi_{gm}$ into a one-halo term, $\xi_{gm}^{1h} =
\bar\delta(r)$, and a two-halo term, $\xi_{gm}^{2h}$, calculated using the
best-fitting $b_0=1.12$ from the previous section.  The one-halo
$\xi_{gm}^{1h}$ differs from the full $\xi_{gm}$ slightly on large scales,
$r\ga 1\hMpc$.  Evidently, there is not much overlap between haloes on the
scales measured, which is not surprising since the high halo mass threshold
precludes a large subhalo fraction in the sample.  However, there is only a
significant signal in $\xi_{gm}$ on scales $r \la 1\hMpc$, limiting the
signal in $\xi_{gm}^{1h}$ and $\xi_{gm}^{2h}$ as well.

What does an `average halo profile' really mean?  In section \ref{mhc}, all
of the halo profiles were identical (up to a multiplicative constant).  In
the real Universe, though, there are haloes of different sizes, shapes, and
environments.  The procedure in the present section gives the average halo
profile under a partition of dark matter (including orphans) into haloes
such that $y_{gm}^{1h} = y_{gm}^{2h}$.  In such a partition, halo profiles
do not depend systematically on the clustering of their galaxies.  Although
this equality holds fairly well in simulations, it cannot hold exactly,
since in the real Universe, both halo profiles and the amplitude of the
galaxy power spectrum depend on halo mass.  However, the question remains:
is it possible to partition the dark matter in the real Universe into
`haloes' around galaxies in a physically meaningful, if somewhat
artificial, way to ensure that $y_{gm}^{1h} = y_{gm}^{2h}$?  If there is,
then our procedure to isolate the $1h$ and $2h$ terms of $\xi_{gm}$ gives
the precise average halo profile if the dark matter is partitioned in this
way.

\section{Conclusion}

This paper introduces a new, galaxy-halo model of large-scale structure.
It is related conceptually to the standard halo model of large-scale
structure, but there are significant differences.  In the standard halo
model, haloes are the fundamental objects; galaxies are placed within them
according to the halo mass.  In the galaxy-halo model, galaxies are the
fundamental objects, which have (galaxy) haloes around them.

One result to come out of the galaxy-halo model is a deeper understanding
of the galaxy-matter cross-correlation coefficient $R$ in terms of a halo
mass dispersion.  Equation (\ref{r2a}), using a few inputs (the galaxy
power spectrum, a large-scale bias, and a dimensionless measure of the
scatter in the halo mass), gives an approximation for $R(k)$ which seems
accurate on mildly non-linear scales, $k \la 3\ihMpc$.  With this
model for $R$, it becomes feasible to measure the galaxy-matter bias
down to scales $k \la 3\ihMpc$ from measurements of the galaxy and
galaxy-matter power spectra (or correlation functions), and thereby to
infer the matter power spectrum down to these scales.

This equation for $R(k)$ has the following intuitive explanation.  On small
scales, the measurement of $R(k)$ samples at most one galaxy at a time.  A
scatter in halo mass thus naturally produces a scatter in the galaxy
density-matter density relationship, producing a small value of $R(k)$
(this value depends on the spread in halo masses).  On large scales, many
haloes are averaged over to measure $R(k)$, reducing the scatter in the
galaxy density-matter density relationship and forcing $R(k)$ toward unity.

The galaxy-halo model also provides a technique for inferring average halo
density profiles, given measurements from a galaxy sample of the galaxy and
galaxy-matter correlation functions.  It is really the one-halo term of the
galaxy-matter correlation function which corresponds to a average halo
density profile; we present and test an algorithm to isolate this term.

Another application of the galaxy-halo model is to the bias $b^2(k) =
P_{gg}(k)/P_{mm}(k)$ between galaxies and matter. If the shot noise is
excluded from the galaxy power spectrum, the bias generally dips down on
intermediate scales where the one-halo and two-halo terms of the matter
power spectrum are comparable.  On small scales, the bias generally
increases with wavenumber because of haloes' extended (not pointlike)
density profiles, which cause a downturn in the matter power spectrum.

\section*{Acknowledgments}
We thank Erin Sheldon and Idit Zehavi for sharing their measurements and
suggestions with us.  We also thank an anonymous referee for suggestions.
This work was supported by NASA ATP award NAG5-10763, NSF grant
AST-0205981, and grants from the National Computational Science Alliance.


\begin{thebibliography}{28}

\bibitem[{{Berlind} \& {Weinberg}}(2002)]{bw}
  Berlind, A.A., Weinberg, D.H., ApJ, 575, 587

\bibitem[{Coles}(1993)]{coles}
  Coles, P., {MNRAS}, 262, 1065

\bibitem[{{Cooray} \& {Sheth}}(2002)]{cs}
  Cooray A., Sheth R., 2002, Phys. Rep., 372, 1

\bibitem[{{Dekel} \& {Lahav}}(1999)]{dl}
  Dekel A., Lahav O., 1999, ApJ, 520, 24
  
\bibitem[{{Guzik} \& {Seljak}}(2002)]{gs02}
  Guzik J., Seljak U., 2002, {MNRAS}, 335, 311
  
\bibitem[{Hamilton}(2000)]{h2000}
  Hamilton A.J.S., 2000, {MNRAS}, 312, 257

\bibitem[{Hamilton}(2005)]{h05}
  Hamilton A.J.S., 2005, to appear in {\it Data Analysis in Cosmology},
  ed.\ V.\ Mart\'{\i}nez, Springer-Verlag Lecture Notes in Physics
  (astro-ph/0503603)
  
\bibitem[{{Hoekstra} et al.}(2003)]{hoek}
  Hoekstra H., van Waerbeke L., Gladders M.D., Mellier Y., Yee H.K.C.,
  2002, ApJ, 577, 604

\bibitem[{{Ma} \& {Fry}}(2000)]{mf}
  Ma C., Fry J.N., 2000, ApJ, 543, 503
  
\bibitem[{{Mandelbaum} et al.}(2005)]{manbaum}
  Mandelbaum R., Tasitsiomi A., Seljak U., Kravtsov A.V., Wechsler R.H., 2005, {MNRAS}, submitted. (astro-ph/0410711)
  
\bibitem[{{Mo} \& {White}}(1996)]{mw}
  Mo, H.J., White, S.D.M., 1996, 282, 347

\bibitem[{{Navarro}, {Frenk} \& {White}}(1997)]{nfw97}
  Navarro J., Frenk C., White S., 1997, ApJ, 490, 493

\bibitem[{{Neyman} \& {Scott}}(1952)]{ney}
  Neyman J., Scott E.L., 1952, ApJ, 116, 144
  
\bibitem[{{Neyrinck}, {Hamilton} \& {Gnedin}}(2004)]{nhg}
  Neyrinck M.C., Hamilton A.J.S., Gnedin N.Y., 2004, {MNRAS}, 341, 1 (NHG)

\bibitem[{{Neyrinck}, {Gnedin} \& {Hamilton}}(2005)]{voboz}
  Neyrinck M.C., Gnedin N.Y., Hamilton A.J.S., 2005, {MNRAS}, 356, 1222

\bibitem[{{Peacock} \& {Smith}}(2000)]{ps}
  Peacock J.A., Smith R.E., 2000, {MNRAS}, 318, 1144
  
\bibitem[{Peebles}(1980)]{peebles}
  Peebles P.J.E., 1980, {\it The Large Scale Structure of the Universe}.
  Princeton Univ.\ Press, Princeton

\bibitem[{Pen}(1998)]{p}
  Pen, U., 1998, ApJ, 504, 601
  
\bibitem[{{Scherrer} \& {Bertschinger}}(1991)]{bert}
  Scherrer R.J., Bertschinger E., 1991, ApJ, 381, 349

\bibitem[{{Scoccimarro} et al.}(2001)]{sshj}
  Scoccimarro R., Sheth R.K., Hui L., Jain B., 2001, ApJ, 546, 20

\bibitem[{Seljak}(2000)]{s00}
  Seljak U., 2000, {MNRAS}, 318, 203

\bibitem[{{Seljak} \& {Warren}(2004)}]{sw}
  Seljak U., Warren M.S., 2004, {MNRAS}, 355, 129
  
\bibitem[{{Seljak} et al.(2005a)}]{s05a}
  Seljak U., et al., 2005a, Phys. Rev. D71, 043511

\bibitem[{{Seljak} et al.(2005b)}]{s05b}
  Seljak U., et al., 2005b, Phys. Rev. D71, 103515
  
\bibitem[{{Sheldon et al.}(2004)}]{sheldon}
  Sheldon E.S. et al., 2004, AJ, 127, 2544

\bibitem[{{Sheth} \& {Tormen}(1999)}]{st99}
  Sheth R.K., Tormen G., 1999, {MNRAS}, 308, 119
  
\bibitem[{{Tasitsiomi et al.}(2004)}]{iro}
  Tasitsiomi A., Kravtsov A.V., Wechsler R.H., Primack J.R., 2004, ApJ, 614, 533

\bibitem[{{Tegmark} \& {Bromley}(1999)}]{teg}
  Tegmark M., Bromley B.C., 1999, ApJ, 518, L69

\bibitem[{{Taruya} \& {Soda}(1999)}]{soda}
  Taruya A., Soda, J., 1999, ApJ, 522, 46
  
\bibitem[{{Zehavi et al.}(2002)}]{z02}
  Zehavi I. et al., 2002, ApJ, 571, 172
  
\end{thebibliography}
\end{document}